%% file: main.tex
\documentclass[sigconf,letter]{acmart}

\usepackage{booktabs} 
\usepackage{float}
\usepackage{subfigure}
\usepackage{bm}
\usepackage{amsmath}
\usepackage{amsthm}
\usepackage{amssymb}
\usepackage{algorithm}
\usepackage{algorithmic}
\usepackage{bbm}
\usepackage[framemethod=1]{mdframed}

\ifodd0
    \newcommand\rev[1]{{\color{red}#1}}
    \newcommand{\com}[1]{\textbf{\color{blue} (COMMENT: #1)}} 
\else
    \newcommand\rev[1]{{#1}}
    \newcommand{\com}[1]{}
\fi

\def \ISTR {}

\theoremstyle{plain}
\newtheorem{thm}{\protect\theoremname}
\theoremstyle{plain}
\newtheorem{defn}[thm]{\protect\definitionname}
\theoremstyle{plain}
\newtheorem{prop}[thm]{\protect\propname}
\theoremstyle{plain}

\theoremstyle{plain}
\newtheorem{corol}[thm]{\protect\corolname}

\providecommand{\definitionname}{Definition}
\providecommand{\propname}{Proposition}
\providecommand{\lemmaname}{Lemma}
\providecommand{\theoremname}{Theorem}
\providecommand{\corolname}{Corollary}

\copyrightyear{2019}
\acmYear{2019}
\setcopyright{acmcopyright}
\acmConference[Mobihoc '19]{The Twentieth ACM International Symposium on Mobile Ad Hoc Networking and Computing}{July 2--5, 2019}{Catania, Italy}
\acmBooktitle{The Twentieth ACM International Symposium on Mobile Ad Hoc Networking and Computing (Mobihoc '19), July 2--5, 2019, Catania, Italy}
\acmPrice{15.00}
\acmDOI{10.1145/3323679.3326512}
\acmISBN{978-1-4503-6764-6/19/07}

\begin{document}
\title{A Probabilistic Approach for Demand-Aware Ride-Sharing Optimization}

\author{Qiulin Lin, Wenjie Xu, Minghua Chen}
\affiliation{
	\department{Information Engineering}
	\institution{The Chinese University of Hong Kong}
	}
\author{Xiaojun Lin}
\affiliation{
	\department{Electrical and Computer Engineering}
	\institution{Purdue University}
	}

\renewcommand{\shortauthors}{Q. Lin, W. Xu, M. Chen, and X. Lin}

\input{abstract.tex}


%
%




\settopmatter{printfolios=true}
\maketitle

\input{intro.tex}

\input{model.tex}
\input{formulation.tex}

\input{algorithm.tex}
\input{simulation.tex}
\input{conclusion.tex}

\section*{Acknowledgement} 
Xiaojun Lin would like to thank the support by NSF Grant ECCS-1509536.

\bibliographystyle{acm}
\bibliography{ref} 

\ifdefined \ISTR 
\input{Appendix.tex}
\fi

\end{document}

%% file: abstract.tex
\begin{abstract}
Ride-sharing is a modern urban-mobility paradigm with tremendous potential
in reducing congestion and pollution. Demand-aware design is a promising
avenue for addressing a critical challenge in ride-sharing systems, namely joint 
optimization of request-vehicle assignment and routing for a fleet of vehicles.
In this paper, we develop a \emph{probabilistic} demand-aware framework to tackle
the challenge. We focus on maximizing the expected number of
\rev{passenger pickups, given the probability distributions of future demands.} 
The key idea of our approach is to assign requests to vehicles in a probabilistic manner. It differentiates our work from existing ones and allows us to explore a richer design space to tackle the request-vehicle assignment puzzle with \rev{a} performance guarantee but still keeping the final solution practically implementable.
The optimization problem is non-convex, combinatorial, and NP-hard in nature. As a key contribution,
we explore the problem structure and propose an elegant approximation of the
objective function to develop a dual-subgradient heuristic. We characterize a condition
under which the heuristic generates a $\left(1-1/e\right)$
approximation solution. Our solution is simple and scalable, amendable for 
practical implementation. Results of numerical experiments based on \rev{real-world traces} in Manhattan show
that, as compared to a conventional demand-oblivious scheme, our demand-aware solution improves the passenger pickups by up to \rev{46\%}. 
The results also show that joint optimization at the fleet level \rev{leads to 19\% more pickups than that by separate optimizations at individual vehicles}.
\end{abstract}

%% file: intro.tex
\section{Introduction \label{sec:intro}}

Dynamic ride-sharing or ride-sharing in short, is a modern paradigm for urban mobility, where passengers with
similar itineraries and time schedules share riders on short-notice. Popular ride-sharing services,
such as uberPOOL\footnote{UberPool. https://www.uber.com/ride/uberpool/} and Lyftline\footnote{LyftLine. https://www.lyft.com/line}, not
only can provide convenient and cost-effective transportation to individuals,
but also can create significant positive impacts on congestion and
pollution. Take Manhattan as an example. The annual cost of congestion
is more than \$20 billion \cite{PFNYC}, which includes 24 million
hours of time lost to sitting in traffic and an extra 500 million
gallons of fuel burned. With ride-sharing, the authors in~\cite{Alonso-Mora17012017,Miller-predict2017}
show that 98\% of the Manhattan rides currently served by over 13,000
taxis could be served with just 3,000 vehicles of capacity four, with marginal increment
in the trip delay. The aggregate trip
distance, an indicator of commute time and gasoline consumption, can
also be reduced by more than 30\%. Overall, ride-sharing offers a
clear opportunity for alleviating congestion, reducing pollution,
and improving transportation efficiency.

A number of societal and economic issues need to be resolved in order
to capitalize the maximum benefit of ride-sharing \cite{agatz2012optimization,clewlow2017disruptive}.
On the technical front, the holy-grail problem is\emph{ how to jointly
optimize the request-vehicle assignments and routing for a fleet of
vehicles (considering future request-vehicle dynamics)}; \rev{see e.g.,~\cite{Alonso-predict2017} for a discussion}. It is a multi-slot
vehicle pickup-and-delivery problem with ride-sharing in consideration, which
is challenging to solve \rev{as} even its single-slot version
is already NP-hard \cite{Alonso-Mora17012017,pillac2013review}.

There are mainly three lines of studies in the literature \cite{furuhata2013ridesharing,agatz2012optimization}.
The first is to devise offline solutions, assuming full knowledge
of future travel requests when making decisions. The offline problem
is known to be NP-hard. The authors in \cite{shengyu} propose a 2.5-approximation
algorithm under a constrained setting. Many studies consider efficient heuristics
and metaheuristics \cite{furuhata2013ridesharing,Ma2013Tshare,Santi16092014,Alonso-Mora17012017,biswas2017profit,zhang2014carpooling,jia2017optimization}.
These offline solutions may serve as performance benchmarks, but they
are usually not practical. The second is to design demand-oblivious
solutions, assuming zero knowledge of future requests when making
decisions; see e.g., \cite{Alonso-Mora17012017,jia2017optimization,Zhang:2013,Ma2013Tshare}.
While demand-oblivious solutions are more amenable for practical implementation,
their performance can be very conservative as they do not adapt to future demand patterns. The third is to develop
\emph{demand-aware} solutions, assuming only distributional information
on future travel requests when making decisions. Thanks to the advance
in machine learning and data analytics, statistical knowledge of future
travel requests can be efficiently learned by leveraging their regular
hourly/daily/weekly patterns \cite{Yuan2011where}. This approach
opens up new design space for optimizing ride-sharing systems, and
the initial success of developing demand-aware ride-sharing routing
solution \cite{QiulinRideSharingRouting17} is encouraging. 

In this paper, we adopt the demand-aware mindset and develop a joint
request-vehicle assignment and routing solution given the distributional
information on future travel requests. A key idea in our design is
to assign requests to vehicles in a \emph{probabilistic} manner. This
allows us to explore a richer design space to tackle the request-to-vehicle
assignment puzzle with performance guarantee, but still keeping the
final solution practically implementable.\footnote{We remark that our probabilistic approach is different from the randomization
mechanism commonly used in algorithm design. Specifically, the joint
assignment and routing solution derived by our approach is not
a time-shared version of multiple optimal deterministic solutions
for different sample-path realizations of the future travel requests.} Our particular study in this paper focuses on maximizing the expected number of
new (ride-sharing) passengers picked up for a fleet of vehicles of
capacity two,
with passenger waiting time limit and passenger transportation deadline taken into account. The problem is important for enhancing the quality of service of
the ride-sharing service platform such as uberPOOL
and Lyftline. It is also equivalent to maximizing
the revenue of VIA\footnote{https://ridewithvia.com/. In VIA, the income of picking up a new passenger is a fixed amount
regardless of the trip distance. The expected total revenue is thus
proportional to the expected number of passengers picked-up.}. Our probabilistic approach is general and can be applied to optimize
other system objectives. Our main contributions in this paper are
as follows.

$\mbox{\ensuremath{\rhd}}$ In Sec.~\ref{sec:system-model}, we propose
a general probabilistic framework for demand-aware ride-sharing optimization.
The framework allows us to explore a bigger design space of joint
request-vehicle assignment and routing with request
statistics taken into account. 

$\mbox{\ensuremath{\rhd}}$ In Sec.~\ref{sec:problem-formulation},
applying the probabilistic framework, we formulate the important problem
of joint request-vehicle assignment and routing for a fleet of vehicles
given request statistics, in order to maximize the expected
number of new (ride-sharing) passengers picked up. The problem is
nonlinear and combinatorial, and we show that it is NP-hard. We then
reformulate the problem into a linear-combinatorial one. We show that
solving the reformulated problem gives an approximation solution to 
the original problem with an approximation ratio of $1-1/e$. 

$\mbox{\ensuremath{\rhd}}$ In Sec.~\ref{sec:algorithm}, the reformulated
linear-combinatorial problem is still challenging, especially for
large-scale instances; indeed, it is still NP-hard. To this end, by
leveraging elegant insights from studying the dual of the re-formulated
problem, we design a scalable heuristic solution. We further characterize a condition under which the heuristic
generates an optimal solution to the reformulated problem, and hence
a $\left(1-1/e\right)$ approximation solution to the original problem.

$\mbox{\ensuremath{\rhd}}$ In Sec.~\ref{sec:simulation}, we carry out numerical experiments based
on real-world travel request traces in Manhattan. The results show
that as compared to a conventional demand-oblivious scheme, our demand-aware solution improves the total number of passenger pickups by up to \rev{46\%. The results also show that joint optimization at the fleet level gives 19\% more pickups than that obtained by individual vehicles carrying out optimization separately. }

\ifx \ISTR \undefined
Due to the space limitation, all proofs are included in our technical
report \cite{TechnicalReport}, unless stated otherwise.
\fi

%% file: model.tex
\begin{figure}[t]
\subfigure[Transportation network $\mathcal{G}_{0}$.\label{fig:T-network.example}]{\includegraphics[width=0.9\columnwidth]{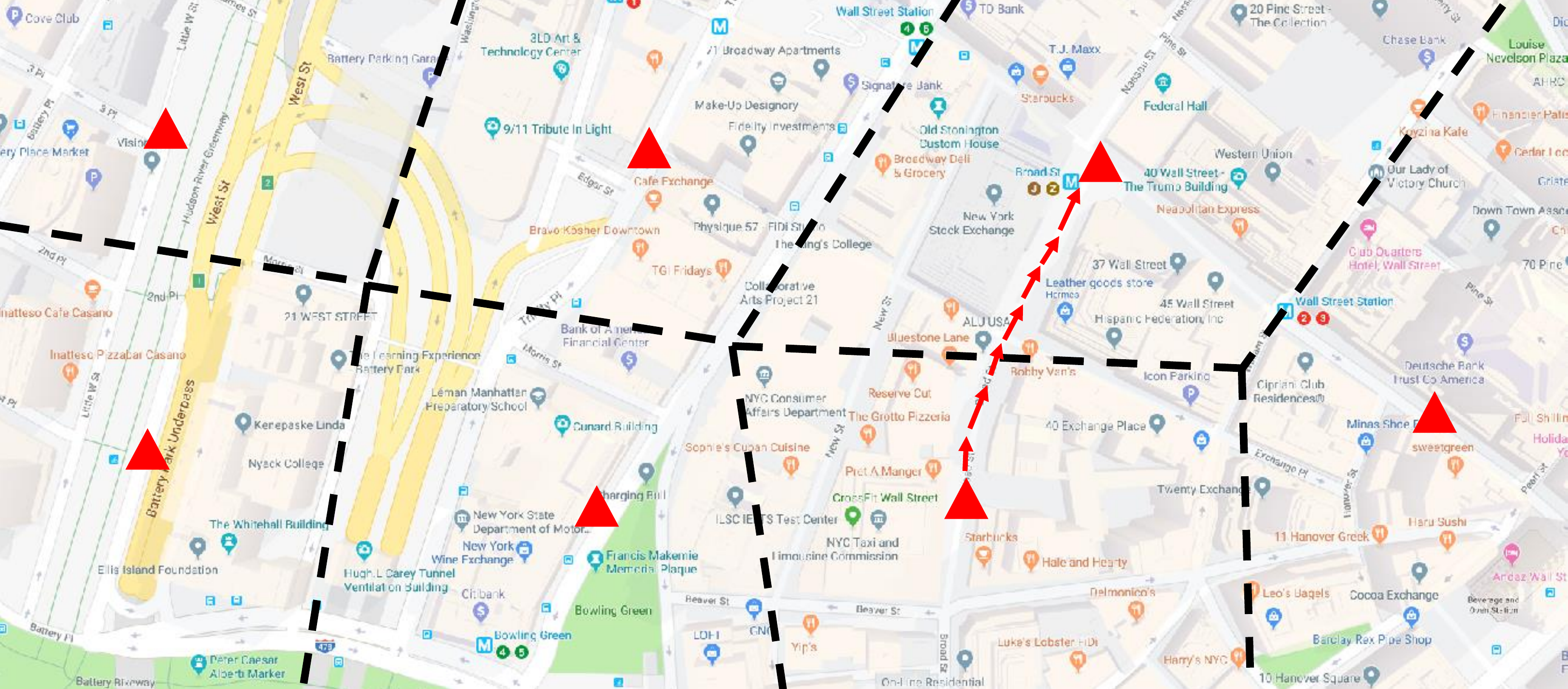}}
\subfigure[Region graph $\mathcal{G}$.\label{fig:R-network.example}]{\includegraphics[width=0.8\columnwidth]{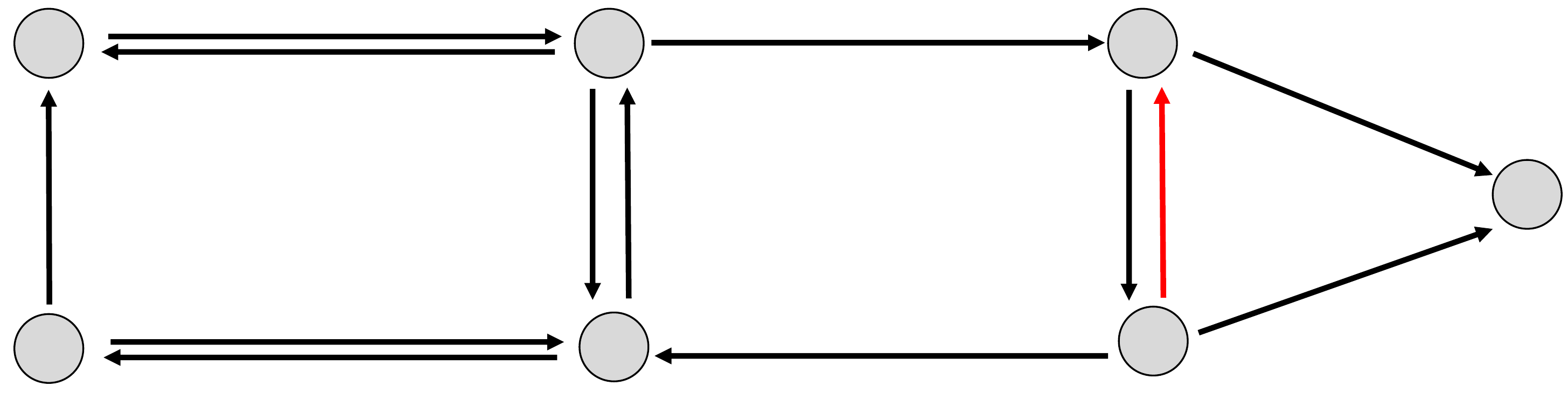}}

\caption{
An example of the transportation network and the corresponding region
graph. 
Each region has a representative node marked as a red triangle. The constructed
region graph \rev{is} shown in Fig.~\ref{fig:R-network.example}. Each
node in the region graph represents a region. Each edge $(u,v)$ in
the region graph represents a fastest path in the transportation network
from the representative node of region $u$ to that of region $v$.
For example, the edge in color red in the region graph in Fig.~\ref{fig:R-network.example}
represents the path in color red in the transportation network in
Fig.~\ref{fig:T-network.example}. \label{fig:region-graph}}
\end{figure}

\section{Problem Settings} \label{sec:system-model}
\rev{ Time is divided into slots of equal length and $\mathcal{T}$
is the set of the slots. We consider the scenario of a fleet of $N$
vehicles of capacity two serving an urban area. We present the system modeling in this section and the problem formulation in the next section.  }

\subsection{Transportation Network and Region Graph \label{ssec:network-model}}

We model the urban transportation network as a directed graph $\mathcal{G}_{0}\triangleq(\mathcal{V}_{0},\mathcal{E}_{0})$
with node set $\mathcal{V}_{0}$ and edge set $\mathcal{E}_{0}$,
as shown in Fig.~\ref{fig:region-graph}. Each edge
$(u_{0},v_{0})\in\mathcal{E}_{0}$ is a road segment from node $u_{0}\in\mathcal{V}_{0}$
to node $v_{0}\in\mathcal{V}_{0}$, and the travel time of edge $(u_{0},v_{0})$
is denoted as $\xi_{u_{0},v_{0}}$ (unit: slots). To introduce our
travel request model later, we construct a region graph $\mathcal{G}\triangleq(\mathcal{V},\mathcal{E})$
with node set $\mathcal{V}$ and edge set $\mathcal{E}$, as shown
in Fig.~\ref{fig:T-network.example}. In particular, we partition
the transportation network $\mathcal{G}_{0}$ into multiple regions. Each node $v\in\mathcal{V}$ in the region graph $\mathcal{G}$
represents a region in $\mathcal{G}_{0}$. We assign a representative
node in each region, from which all other nodes in the region can
be reached in a small number of slots, e.g., 5 slots, illustrated
as the black dots in Fig.~\ref{fig:R-network.example}\footnote{For a general urban road network, constructing the regions is equivalent
to solving a clustering problem to find a set of clusters. Location
points within each cluster are close to each other. The problem can
be solved by using celebrated algorithms like k-means \cite{mackay2003information}.
For a dense and regular urban road network like Manhattan, one can
simply partition the district into regions of equal area. We adopt
this method in the simulation in Sec.~\ref{sec:simulation}.}. We add a directed edge $(u,v)$ in $\mathcal{G}$, as illustrated
in \ref{fig:T-network.example}, if there exists a
path from the representative node of region $u$ to that of region
$v$ in the road graph such that the major fraction of the path is
within those two regions\footnote{More specifically, for any two nodes $u,v\in\mathcal{V}$, we denote
$T(u,v)$ as the minimal travel time from the representative node
of region $u$ to that of region $v$ in the transportation network.
Then there is an edge from $u$ to $v$ in the region graph if $T(u,v)\le\eta\cdot(T(u,k)+T(k,v))$
for any region $k\in\mathcal{G}$. In our simulation in Sec.~\ref{sec:simulation},
we set $\eta=0.8$.}. Let $\delta_{u,v}$ be the travel time of the fastest path.

\rev{
We assume that when a travel request appears in a region, it only
waits for a limited amount of time, e.g.,5 slots. As
such, only the vehicles in the same region can pick up the passenger.
This captures the observation that each passenger is associated with
a waiting-time limit, e.g., 10 minutes for uberPOOL; vehicles  outside the region cannot pick up the passenger on time.
Hence, the ``size'' of regions can be determined by the waiting-time
limit set by the ride-sharing system. Our road network and region
graph models are similar to those used in the literature; see e.g.,
\cite{QiulinRideSharingRouting17,EmptyCar,braverman2016empty}. All our modeling and analysis are based on the region graph $\mathcal{G}$.}

\subsection{Travel Request \label{ssec:demand-model}}

Each travel request consists of the time of request, a pickup region
$v\in\mathcal{V}$ and a drop-off region $u\in\mathcal{V}$, a waiting
time limit, and a trip deadline. We assume that the waiting-time limit
for all travel requests to be the same, e.g., 5 slots, and it is used
to properly gauge the size of regions as described in Sec. \ref{ssec:network-model}.
The trip deadline for a travel request from $v$ to $u$ is denoted
as $\Delta_{v,u}^{\textsf{max}}$ slots, and we note that UberPOOL
has already provided such ``Arrive By'' service\footnote{uberPOOL Just Got More Punctual. https://newsroom.uber.com/punctual-uberpool/.}.
In our study, we set the $\Delta_{v,u}^{\textsf{max}}=\alpha\cdot\delta_{v,u}$,
where $\alpha>1$ represents the delay tolerance factor and $\delta_{v,u}$
denotes the travel time of the fastest path from the representative node in region $v$ to that in region $u$. 

Under the \emph{demand-aware} setting, the time-dependent distributional
information on (future) travel requests are available. Specifically,
the probability that at time slot $t$, there are $k$ passengers
appearing at region $v$ on $\mathcal{G}$ and going to region $u$
is given as $p_{v,u,k}^{t}$. Without loss of practical relevance,
we assume that there are at most $K>0$ passengers going from $v$
to $u$ at any given time. Apparently, $\sum_{k=1}^{K}p_{v,u,k}^{t}=1,\;\forall u\in\mathcal{V}.$
We further assume that travel request arrivals across different nodes and in different time slots are independent to each other. 

\subsection{Vehicle State}
At a given time $t$, each of the $N$ vehicles
is in one of the three states: 
\begin{itemize}
    \item (i) the vehicle is delivering two
passengers on board and will not pick up any passenger,
    \item (ii) the
vehicle is delivering one passenger on board and can pick up one more
ride-sharing passenger,
    \item  \rev{and (iii) the vehicle is empty and roaming
towards a pre-selected hot spot (e.g., regions with good chances of picking up new passengers), with a self-selected and usually sufficiently large deadline. }
\end{itemize} 
\rev{Note that an empty vehicle in state (iii) can be regarded as a vehicle in state (ii) with one ``virtual'' passenger on board and
is looking for picking up a new passenger along the way to the hot spot. Thus, from the modeling point of view, there is no  difference between state (ii) and state (iii). 

We also note that a vehicle may transit between states upon passenger pickup and delivery. For the empty vehicle in state (iii), its state will be updated to (ii) once it  picks up a new passenger. Similarly, a vehicle's status will change from state (ii) to state (iii) if it picks up a ride-sharing passenger. Furthermore, a vehicle's state will change from state (i) or (ii) to state (iii) if the passenger(s) on board are delivered. \com{will be good to include a state transition diagram to make this super easy to understand.}}


\subsection{Request-Vehicle Assignment and Routing\label{ssec:ride-sharing-scenario}}

Under the offline setting, the travel requests are assumed to be known in advanced.
The requests are assigned to vehicles in the same regions by solving
a combinatorial puzzle, taking into account \rev{the vehicle states} and capacities,
the waiting-time limit of the requests, and the system objective.
This step is also called a trip-vehicle matching in the literature.
 Given the assignments, individual vehicles then compute the best
pickup-delivery routes to transport the passengers. 
It is known that
the joint assignment and routing problem is NP-hard and challenging
to solve; \rev{see e.g.,~\cite{shengyu}}. Furthermore, such an offline setting can be impractical because the exact information of future travel requests is
usually not available.

\rev{Under the demand-aware setting, the distributional information
of travel requests is available.} In particular, at time $t$, with probability $p_{v,u,k}^{t}$, there are $k$ travel requests appearing in region $v$ and going to region $u$. We expect that such distributional information is much easier to obtain in practice. \rev{The key idea in our demand-aware design is to assign requests to vehicles in a probabilistic manner. Specifically, let $y_{i,v,u,k}^{t}\in\left[0,p_{v,u,k}^{t}\right]$
be the probability of the joint event that (i) $k$ travel requests
appear in region $v$ going to region $u$ at time $t$ and (ii) one of the requests is assigned to vehicle $i$ ($1\leq i\leq N$). We remark that $y_{i,v,u,k}^{t}$'s are to be designed later, and they need to satisfy a set of conditions to be practically feasible, i.e., there exists a practical scheme that can realize the probabilistic assignment. We will present the feasibility conditions in the next section. Given $y_{i,v,u,k}^{t}$'s, upon $k$ travel requests appearing in region $u$ and going to region $v$ at time $t$, we assign one of the $k$ requests to vehicle $i$ with probability $y_{i,v,u,k}^{t}/p_{v,u,k}^{t}$. The overall probability that vehicle $i$ is assigned a travel request going from region $v$ to $u$ at time $t$ is $\sum_{k=1}^{K}y_{i,v,u,k}^{t}\text{\ensuremath{\in\left[0,1\right]}}$. The probability that vehicle $i$ is assigned a  request in region $v$ at time $t$ (going to any region) is 
\begin{equation}
y_{i,v}^{t}\triangleq1-\prod_{u\in\mathcal{V}}\left(1-\sum_{k=1}^{K}y_{i,v,u,k}^{t}\right),\;\;\forall1\le i\leq N,v\in\mathcal{V},t\in\mathcal{T}.\label{probsum}
\end{equation}

Given the probabilities of picking up (new or ride-sharing) passengers in individual regions in every time epoch, individual vehicles compute their own routes to maximize the expected reward or minimize the expected cost. 
As it will become clearer in the next section, the joint probabilistic assignment
and routing problem admits bigger design space for designing efficient algorithms. 
}

%% file: formulation.tex
\section{Problem Formulation\label{sec:problem-formulation}}
In this section, we formulate the joint probabilistic request-vehicle assignment and vehicle routing problem for the fleet of $N$ vehicles. \rev{Without loss of generality, we assume that we start at time $t=0$ and all vehicles are in state (ii), which also covers vehicles in state (iii) as there is no difference between the two from the modeling perspective.} Our objective is to maximize
the expected number of new or ride-sharing passenger pickups by
the fleet from time $t=0$ to the first time epoch involving a change in the vehicle states, i.e., a passenger arriving at the destination or an empty vehicle picking up a new passenger. To optimize the overall performance in a time horizon, e.g., a day, the optimization will be re-carried out at these state-changing time epochs. 

\subsection{Ride-Sharing Feasibility \label{ssec:ride-sharing.feasibility}}

\begin{defn}
\label{def:ride-sharing.feasibility} Suppose that vehicle $i$ is
transporting a passenger from region $s_{i}$ to region $d_{i}$ and passing through
region $v$ at time $t$. We say that a $s_{i}\rightarrow v\rightarrow\left\{ u,d_{i}\right\} $
ride-sharing plan for vehicle $i$ at time $t$ is feasible if the vehicle
can pick up another passenger from $v$ to $u$ at time $t$ and
deliver both passengers before their trip deadlines.
\end{defn}

Let $\mathcal{R}_{v,u,d_{i}}$ denote the set of two types of routes: (i) going
from $v$ to $u$ and then to $d_{i}$, all by the fastest paths,
and (ii) going from $v$ to $d_{i}$ and then to $u$, all by the
fastest paths. Let $\mathit{z_{i,v,u}^{t}}$  \rev{$\in\{0,1\}$} be the indicator variable
of whether \rev{a \textbf{ }$s_{i}\rightarrow v\rightarrow\left\{ u,d_{i}\right\}$
ride-sharing plan} for the vehicle $i$ at time $t$ is feasible, i.e.,
whether the following problem has a feasible solution: 
\begin{align}
\max\;\; & 1\nonumber \\
\mbox{s.t.}\;\; & \gamma_{v,u}\left(r\right)\leq\Delta_{v,u}^{\max},\label{eq:cons:delay1}\\
 & \gamma_{s_{i},v}+\gamma_{v,d_{i}}\left(r\right)\leq\Delta_{s_{i},d_{i}}^{\max},\label{eq:cons.delay2}\\
\mbox{var.\;\;} & r\in\mathcal{R}_{v,u,d_{i}},\nonumber 
\end{align}
where $\Delta_{v,u}^{\max}$ and $\Delta_{s_{i},d_{i}}^{\max}$ are
trip deadlines, $\gamma_{s_{i},v}$ is the amount of time the vehicle
$i$ already spent in traveling from $s_{i}$ to $v$, and $\gamma_{v,u}(r)$
and $\gamma_{v,d_{i}}(r)$ are the travel times from region $v$ to region
$u$ and region $d_{i}$ along one of the two routes $r$ in $\mathcal{R}_{v,u,d_{i}}$,
respectively. The problem involves finding two fastest paths and some
simple calculus and it is easy to solve. For each vehicle
$i$, we need to solve the problem for every $\left(s_{i},d_{i}\right)$
pair, every $\left(v,u\right)$ pair, and every $\gamma_{s_{i},v}\in\left[0,\Delta_{s_{i},d_{i}}^{\max}\right]$.
The total complexity is polynomial in the size of the region graph and the maximum trip deadline. \rev{We note that these problems can be solved beforehand and the feasibility indicators $\mathit{z_{i,v,u}^{t}}$ and the corresponding feasible paths can be stored for lookup. }

\subsection{Feasibility of Request-Vehicle Assignments\label{ssec:joint.request-vehicle.assignment.and.routing}}

Recall that $p_{v,u,k}^{t}$ is the probability of $k$ travel requests
appearing at time $t$ in region $v$ and going to region $u$. Variable
$y_{i,v,u,k}^{t}$ is the probability of the joint event that (i)
$k$ travel requests appear in region $u$ going to region $v$ at
time $t$ and (ii) one of the requests is assigned to vehicle $i$
($1\leq i\leq N$).  For all $1\leq i\leq N$, define \[
\mathbf{y_{i}}\triangleq\left[y_{i,v,u,k}^{t},\;\forall t\in\mathcal{T},v,u\in\mathcal{V},1\leq k\leq \rev{K}\right].
\]
The following proposition characterizes the feasible region of $\mathbf{y_{i}}$'s. 
\begin{mdframed}[skipabove=10pt, skipbelow=10pt, roundcorner=10pt, linewidth=0pt, backgroundcolor=gray!10] 
\begin{prop}
\label{prop:feasible-region}The request-vehicle assignment probability
vectors $\left[\boldsymbol{y}_{i},1\leq i\leq N\right]$ are feasible if and only if {they satisfy that, for all $1\leq i\leq N$, $1\leq k\leq K$,
$t\in\mathcal{T}$, and $v,u\in\mathcal{V}$, }
\begin{alignat}{1}
y_{i,v,u,k}^{t} & \leq p_{v,u,k}^{t},\label{cons:proballocation2}\\
0\leq y_{i,v,u,k}^{t} & \leq\mathit{z_{i,v,u}^{t}},\label{cons:feasibility}\\
\sum_{i=1}^{N}y_{i,v,u,k}^{t} & \leq k\cdot p_{v,u,k}^{t}.\label{cons:proballocation1}
\end{alignat}
Let $\mathcal{Y}$ be the set of all feasible request-vehicle assignment vectors.
\end{prop}
\end{mdframed}
The inequalities in (\ref{cons:proballocation2}) state that the
assignment probability $y_{i,v,u,k}^{t}$ should not be larger than
$p_{v,u,k}^{t}$, the probability that the $k$ requests appear. The
inequalities in (\ref{cons:feasibility}) state that the assignment
probability can be positive only if the assignment is feasible (see
Definition \ref{def:ride-sharing.feasibility}). The inequalities
in (\ref{cons:proballocation1}) can be understood as follows. First,
$y_{i,v,u,k}^{t}/P_{v,u,k}^{t}$ is simply the \emph{conditional}
probability that given $k$ requests appearing in region $v$ at time
$t$ and going to region $u$, one request is assigned to vehicle
$i$. The inequalities in (\ref{cons:proballocation1}) say
that the conditional expected total number of requests assigned to the fleet of $N$ vehicles should be bounded by $k$. 

Proposition~\ref{prop:feasible-region} lays down an important foundation for our probabilistic demand-award approach. First, it characterizes the necessary and sufficient conditions for a request-vehicle assignment probability vector. Second, it says that there exists a scheme that can realize the probabilistic request-vehicle assignment, such that  the conditional probability of vehicle $i$ being assigned
a passenger out of $k$ appearing requests from $v$ to $u$ at time $t$ is exactly the desired conditional probability $y_{i,v,u,k}^{t}/p_{v,u,k}^{t}$. Specifically, we present one such scheme as follows, which is very similar to the ones in~\cite{ioannidis2018adaptive, blaszczyszyn2015optimal} for network caching system designs. The scheme is also applied for assigning requests to vehicles later in our simulations.

\begin{figure}[ht] 
    \centering 
    \includegraphics[width=0.9\columnwidth]{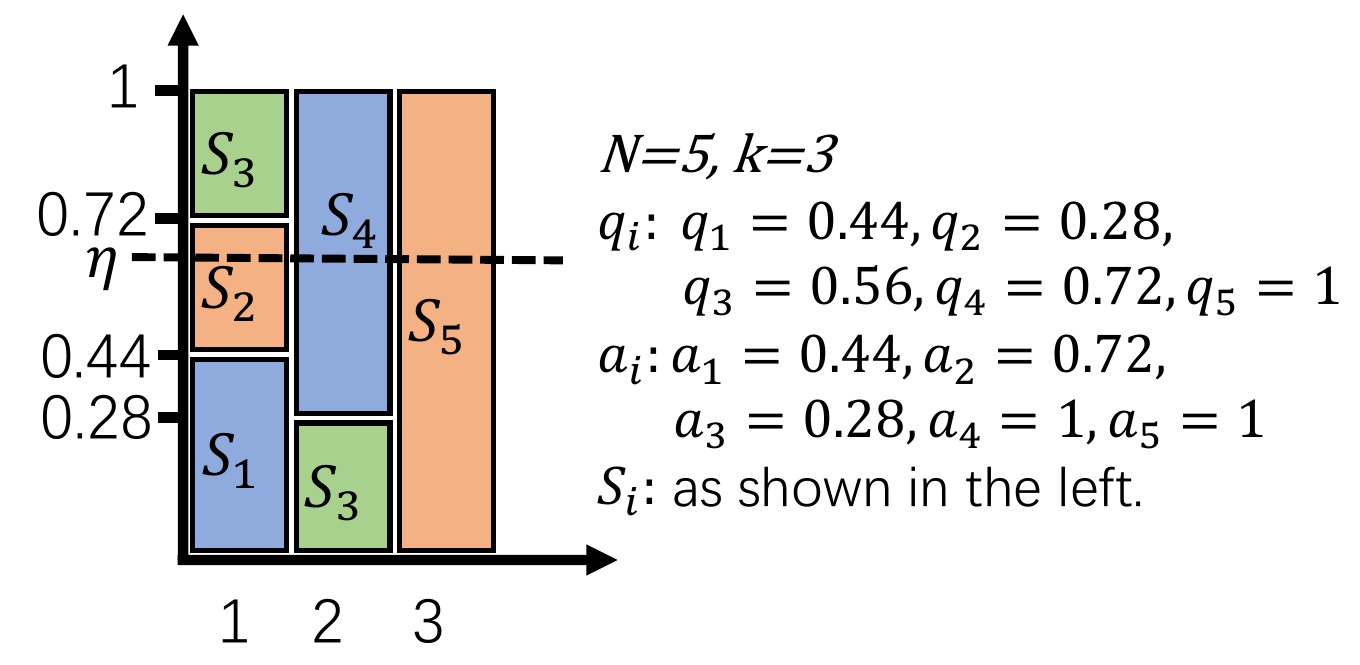}
    \caption{In this particular instance where 3 requests appear in a region of 5 vehicles, the 3 requests are assigned to vehicle 2, 4, and 5. \label{fig:allocation}}
\end{figure}

\textbf{A probabilistic request-vehicle assignment scheme.} Given $k$ appearing requests from $v$ to $u$ at time $t$ and $\left[\boldsymbol{y}_{i},1\leq i\leq N\right]$ satisfying the conditions in \eqref{cons:proballocation2}-\eqref{cons:proballocation1}, we define 
\[q_i = \min\left\{y_{i,v,u,k}^{t}/p_{v,u,k}^{t},\mathit{z_{i,v,u}^{t}}\right\}.
\]
Let $a_0=0$ and $a_i= (a_{i-1}+q_i)-1_{a_{i-1}+q_i>1}$, $1\leq i\leq N$. We associate each vehicle $i$ with an interval $S_i\subset [0,1]$ as follows:
\[S_i=
\begin{cases}
[a_i,a_{i-1}), & a_i > a_{i-1};\\
\emptyset, & a_i = a_{i-1} \text{ and } q_i=0; \\
[0,1],  &a_i = a_{i-1} \text{ and } q_i=1; \\
[a_{i-1},1]\cup[0,a_i),&a_i < a_{i-1}.
\end{cases}
\]
It's straightforward to check that the length of interval $S_i$ is $q_i$.
Then, we generate a number $\eta$ in $[0,1]$ uniformly at random. We assign one of the $k$ requests to vehicle $i$ if $\eta\in S_i$. Since $|S_i|=q_i$, the probability that vehicle $i$ is assigned a request is exactly $q_i$. Also for any value of $\eta$, at most $k$ vehicles are assigned requests. An illustrating example is shown in Fig.~\ref{fig:allocation}. 

\subsection{Problem Formulation}
Suppose vehicle $i$ transports passenger following the path
\[
r_{i}=\left(s_{i},{v}_{i,1},{v}_{i,2},\cdots\cdots,{v}_{i,n_{i}},d_{i}\right)\in\mathcal{R}_{i},
\]
where $\mathcal{R}_{i}$ is the set of all the routes from $s_{i}$ to $d_{i}$ with travel time smaller than $\Delta_{s_i,d_i}^{\max}$ and $n_{i}$
is the number of intermediate regions in route $r_{i}$. 
Given the probabilities of getting
request assignment, i.e., $\mathbf{y_{i}}$, the expected number of passengers that vehicle $i$ will pick up (and starts a feasible ride-sharing
trip) is given by 
\begin{align}
f_{i}\left(\mathbb{\mathbf{y_{i}}},r_{i}\right)\triangleq & 1-\prod_{j=1}^{n_{i}}\left(1-y_{i,{v}_{i,j}}^{t_{i,j}}\right)\nonumber \\
= & 1-\prod_{j=1}^{n_{i}}\prod_{u\in\mathcal{V}}\left(1-\sum_{k=1}^{K}y_{i,{v}_{i,j},u,k}^{t_{i,j}}\right),\label{eq:vehicle.pick-up.prob}
\end{align}
where $\left(1-y_{i,{v}_{i,j}}^{t_{i,j}}\right)$ is the probability
that the vehicle $i$ will not pick up a passenger when passing region
${v}_{i,j}$ at time $t_{i, j}$ and hence $f_{i}\left(\mathbb{\mathbf{y_{i}}},r_{i}\right)$
is the probability that vehicle $i$ will pick up a passenger along
the route $r_{i}$. Since all vehicles have already picked up a passenger
when taking their routes $r_{i}$, each vehicle has only one seat
and can at most pick up one more passenger. As such,
$f_{i}\left(\mathbb{\mathbf{y_{i}}},r_{i}\right)$ is also the expected
number of passengers that vehicle $i$ will pick up along the route
$r_{i}$. We note that $f_{i}\left(\mathbb{\mathbf{y_{i}}},r_{i}\right)$
is non-convex in $\mathbb{\mathbf{y_{i}}}$ and it involves combinatorial routing decisions $r_{i}$'s.

We formulate the joint (probabilistic) request-vehicle assignment and vehicle routing
problem as follows:
\begin{eqnarray}
\mbox{\textbf{MP}}: & \max & \sum_{i=1}^{N}f_{i}\left(\mathbf{y_{i}},r_{i}\right)\label{Prob_MP:obj}\\
 & \mbox{var.} & \left[\boldsymbol{y}_{i},1\leq i\leq N\right]\in\mathcal{Y},\;r_{i}\in\mathcal{R}_{i},\;1\leq i\leq N,\nonumber 
\end{eqnarray}
where $\mathcal{Y}$ is the feasible set of $\boldsymbol{y}_{i}$'s defined in Proposition~\ref{prop:feasible-region}. The following proposition shows that Problem \textbf{MP} is NP-hard. 
\begin{mdframed}[skipabove=10pt, skipbelow=10pt, roundcorner=10pt, linewidth=0pt, backgroundcolor=gray!10]
\begin{prop}
\label{prop:MP-is-NP-hard}The problem \textbf{MP }is NP-hard as it covers the NP-hard single-vehicle demand-aware routing problem in \cite{QiulinRideSharingRouting17}
as a special case. 
\end{prop}
\end{mdframed}

%% file: algorithm.tex
\section{A Dual-Subgradient Algorithm}\label{sec:algorithm}
In this section, we first introduce a time-expanded graph and then study a linear-combinatorial problem,
by leveraging on an approximation for $f_{i}\left(\mathbb{\mathbf{y_{i}}},r_{i}\right)$
and an elegant reformulation. We will then explore the insights
from studying the dual of the linear-combinatorial problem to derive a dual-subgradient algorithm for the original problem \textbf{MP}.

\subsection{Reformulation over a Time-Expanded Graph}
To facilitate the discussions and problem re-formulation for algorithm design, we first construct a time-expanded graph as follows.
\rev{
\begin{defn}
Given $\tau=\max_{1\leq i\leq N}\{\Delta_{s_i,d_i}^{\max}\}$ and the regional graph $\mathcal{G}=(\mathcal{V},\mathcal{E})$,
the time-expanded graph $\mathcal{G}^{[\tau]}=\left(\mathcal{V}^{[\tau]},\mathcal{E}^{[\tau]}\right)$
contains 
\begin{itemize}
    \item $(1+\tau)\times|\mathcal{V}|$ nodes, labeled as
$v^{t}$ where $v\in\mathcal{V}$ and $ t\in[0,\tau]$.

    \item $N$ virtual destination nodes, labeled as $d_i^{-i}$, when $1 \leq i\leq N$ and $d_i$ is the destination of the current rider at vehicle $i$.
\end{itemize}
The edge set $\mathcal{E}^{[\tau]}$
is constructed as follows: 
\begin{itemize}
    \item For each edge $e=(u,v)\in\mathcal{E}$
with travel delay $\delta_{u,v}$, for each $t\in[1,\tau-\delta_{u,v}]$,
create an edge $e^{t}\in\mathcal{E}^{[\tau]}$ from $u^{t}$
to $v^{t+\delta_{u,v}}$. Note that there is no travel delay associated
with $e^{t}$
    \item for each vehicle $i$, for each $t\in[1:\Delta_{s_i,d_i}^{\max}]$, create an edge from $d_i^t$ to $d_i^{-i}$.
\end{itemize}
\end{defn}
}
For any node $v^{t}\in\mathcal{V}^{[\tau]}$, it is associated with the probabilities $\left[p_{v,u,k}^{t}, \forall u\in\mathcal{V}, 1\leq k\leq K\right]$.
For ease of expression, in the following ,we use $\bar{v}$ to represent
the nodes in $\mathcal{V}^{[\tau]}$; note that each $\bar{v}$ is one-to-one
map to a $v^{t}$ is associated with the time $t$ implicitly. Similarly, we use $y_{i,\bar{v},u,k}$ to represent $y^t_{i,v,u,k}$ and $y_{i,\bar{v}}$ to represent $y^t_{i,v}$. We also use  $r_i$ and $f_i(\cdot,\cdot)$ to represent the route of vehicle $i$ and its probability of picking up along route $r_i$, over the time-expanded graph.


Let $\mathcal{R}_{i}^{[\tau]}$ be the set of all the routes from $s_{i}^{0}$ to $d_{i}^{-i}$ on the time expanded graph, 
we reformulate the problem \textbf{MP} over the time-expanded graph as follows:
\begin{eqnarray}
\mbox{\textbf{MP-T}}: & \max & \sum_{i=1}^{N}f_{i}\left(\mathbf{y_{i}},r_{i}\right)\label{Prob_MPT:obj}\\
 & \mbox{var.} & \left[\boldsymbol{y}_{i},1\leq i\leq N\right]\in\mathcal{Y},\;r_{i}\in\mathcal{R}_{i}^{[\tau]},\;1\leq i\leq N,\nonumber 
\end{eqnarray}

\rev{By introducing the time-expanded graph, we omit the delay constraint on each route $r_i$ as ${R}_{i}^{[\tau]}$ is all the simple path (i.e., involving no loop) from $s^0$ to $d_i^{-i}$ in the time-expanded graph. However, we note that the network size of the time-expanded graph become $\tau$ times of the original graph. In the case that all the probabilities are time-invariant, this actually leads to exponential increment of the input size  as it is polynomial in $\tau$ which is exponential in the bit length of the input $\tau$. Consequently, any polynomial time algorithm on the time-expanded graph becomes a pseudo-polynomial time algorithm to the original problem \textbf{MP}}.

\subsection{{A $\left(1-1/e\right)$ Approximation}} 
We first note that $f_{i}\left(\boldsymbol{y}_{i},r_{i}\right)$ in
(\ref{Prob_MPT:obj}) can be upper-bounded and lower-bounded
by two simple concave functions. 
\begin{mdframed}[nobreak, skipabove=10pt, skipbelow=10pt, roundcorner=10pt, linewidth=0pt, backgroundcolor=gray!10] 
\begin{prop}
\label{prop:upper-lower-bounds}Define a concave function 
\[
g_{i}\left(\mathbf{y_{i}},r_{i}\right)\triangleq\min\left(1,\sum_{j=1}^{n_{i}}\sum_{u\in\mathcal{V}}\sum_{k=1}^{K}y_{i,\bar{v}_{i,j},u,k}\right).
\]
Then 
\[
\left(1-1/e\right)g_{i}\left(\mathbf{y_{i}},r_{i}\right)\leq f_{i}\left(\boldsymbol{y}_{i},r_{i}\right)\leq g_{i}\left(\mathbf{y_{i}},r_{i}\right).
\]
\end{prop}
\end{mdframed}
The upper bound is obtained by {the standard union-bound argument} and the lower bound is according to \cite{goemans1994new}. 

With the above understanding, we formulate a problem \textbf{MP-A}
as follows:
\begin{eqnarray}
\text{\mbox{\textbf{MP-A}}}: & \max & \sum_{i=1}^{N}g_{i}\left(\mathbf{y_{i}},r_{i}\right)\label{Prob_MP-A:obj}\\
 & \mbox{var.} & \left[\boldsymbol{y}_{i},1\leq i\leq N\right]\in\mathcal{Y},\;r_{i}\in\mathcal{R}_{i}^{[\tau]},\;1\leq i\leq N.\nonumber 
\end{eqnarray}
It has the same feasible region as \textbf{MP-T} but the objective function
is replaced by a concave one. The following theorem says that interestingly,
solving \textbf{MP-A} gives an approximation solution to \textbf{MP-T}. 
\begin{mdframed}[skipabove=10pt, skipbelow=10pt, roundcorner=10pt, linewidth=0pt, backgroundcolor=gray!10] 
\begin{thm}
\label{thm:approx}Let $\left(\boldsymbol{y}_{i}^{*},r_{i}^{*}\right)$,
$1\leq i\leq N$, be an optimal solution to \textbf{MP-T}, and let $\left(\bar{\boldsymbol{y}}_{i},\bar{r}_{i}\right)$,
$1\leq i\leq N$, be that of \textbf{MP-A}. Then 
\[
\left(1-1/e\right)\sum_{i=1}^{N}f_{i}\left(\boldsymbol{y}_{i}^{*},r_{i}^{*}\right)\leq\sum_{i=1}^{N}f_{i}\left(\bar{\boldsymbol{y}}_{i},\bar{r}_{i}\right)\leq\sum_{i=1}^{N}f_{i}\left(\boldsymbol{y}_{i}^{*},r_{i}^{*}\right).
\]
\end{thm}
\end{mdframed}
\begin{proof}
By the definition of $\left(\boldsymbol{y}_{i}^{*},r_{i}^{*}\right)$,
we have $\sum_{i=1}^{N}f_{i}\left(\bar{\boldsymbol{y}}_{i},\bar{r}_{i}\right)\leq\sum_{i=1}^{N}f_{i}\left(\boldsymbol{y}_{i}^{*},r_{i}^{*}\right)$.
We also have

\begin{align*}
\left(1-1/e\right)\sum_{i=1}^{N}f_{i}\left(\boldsymbol{y}_{i}^{*},r_{i}^{*}\right)\leq & \left(1-1/e\right)\sum_{i=1}^{N}g_{i}\left(\boldsymbol{y}_{i}^{*},r_{i}^{*}\right)\\
\leq & \left(1-1/e\right)\sum_{i=1}^{N}g_{i}\left(\bar{\boldsymbol{y}}_{i},\bar{r}_{i}\right)\\
\leq & \sum_{i=1}^{N}f_{i}\left(\bar{\boldsymbol{y}}_{i},\bar{r}_{i}\right),
\end{align*}
where the first and third steps utilize Proposition \ref{prop:upper-lower-bounds},
and the second step uses the definition of $\left(\bar{\boldsymbol{y}}_{i},\bar{r}_{i}\right)$. 
\end{proof}
An important insight from Theorem \ref{thm:approx} is that the
optimal solution of \textbf{MP-A} is a $\left(1-1/e\right)$ approximation
solution of \textbf{MP-T}. 

\subsection{A Linear-Combinatorial Reformulation} 

We present an \emph{equivalent}
formulation of \textbf{MP-A} to facilitate the algorithm design discussion
later. To proceed, we first introduce a set of routing variables $\mathbf{x}_{i}=[x_{i,\bar{v}}]_{\bar{v}\in\mathcal{V}^{[\tau]}}$ to indicate whether vehicle $i$ pass node $\bar{v}$ on the time-expanded regional graph $\mathcal{R}_{i}^{[\tau]}$:
\begin{equation}
\label{eq:route2node}
    x_{i,\bar{v}}=\begin{cases}
		1, & \text{if the route $r_i$ passes through $\bar{v}$}; \\
		0, & \text{otherwise}.
	\end{cases}
\end{equation}
Define $\mathcal{X}_{i}$ as the set of all valid $\mathbf{x}_{i}$ that corresponds to a $r_{i}\in\mathcal{R}_{i}^{[\tau]}$. It should be clear that the time-expanded graph is acyclic and directed, and any valid $\mathbf{x}_{i}\in \mathcal{X}_{i}$ maps to a valid $r_{i}\in\mathcal{R}_{i}^{[\tau]}$ and vice verse. 

Next, we observe that for $[\mathbf{y}_i, 1\leq i \leq N] \in \mathcal{Y}$, we can replace $\mathcal{Y}$ by $\bar{\mathcal{Y}}$, which is the set of $[\mathbf{y}_i, 1\leq i \leq N]$ that satisfies
\begin{eqnarray}
 &  & \sum_{i=1}^{N}x_{i,\bar{v}}\cdot y_{i,\bar{v},u,k}\leq k\cdot p_{\bar{v},u,k},\label{cons:allocation_x}, \forall \bar{v},u,k\\
 &  & 0\leq y_{i,\bar{v},u,k}\leq p_{\bar{v},u,k}\cdot \mathit{z_{i,\bar{v},u}},\forall i,\bar{v},u,k \label{cons:feasibility_x}
\end{eqnarray}
where $\mathit{z_{i,\bar{v},u}}$ is defined in Sec.~\ref{ssec:ride-sharing.feasibility} and (\ref{cons:feasibility_x}) is equivalent to (\ref{cons:feasibility}) and (\ref{cons:proballocation2}). {As compared to} (\ref{cons:proballocation1}), we only count the probability allocation to vehicles that pass node $\bar{v}$ in (\ref{cons:allocation_x}). Although $\bar{\mathcal{Y}}$ is {larger than $\bar{\mathcal{Y}}$, we can easily see that any feasible solution in $\bar{\mathcal{Y}}$ can map to a feasible solution in $\mathcal{Y}$. Hence, we replace the constraints $[\mathbf{y}_i, 1\leq i \leq N] \in \mathcal{Y}$ by $[\mathbf{y}_i, 1\leq i \leq N] \in \bar{\mathcal{Y}}$.}

Finally, we replace (\ref{cons:feasibility_x}) with $0\leq x_{i,\bar{v}}\cdot y_{i,\bar{v},u,k}\leq x_{i,\bar{v}}\cdot p_{\bar{v},u,k}\cdot\mathit{z_{i,\bar{v},u}}$, introduce new variables $y'_{i,\bar{v},u,k}=x_{i.\bar{v}}\cdot y_{i,\bar{v},u,k}$, and reformulate the problem \textbf{MP-A} as follows:
\begin{align}
\mbox{\textbf{MP-AN}}: \max \;\;& \sum_{i=1}^{N}\sum_{\bar{v}\in\mathcal{V}^{[\tau]}}\left(\sum_{u\in\mathcal{V}}\sum_{k=1}^{K}y'_{i,\bar{v},u,k}\right)\label{obj:MP-N2}\\
  s.t. \;\;& \sum_{\bar{v}\in\mathcal{V}^{[\tau]}}\left(\sum_{u\in\mathcal{V}}\sum_{k=1}^{K}y'_{i,\bar{v},u,k}\right)\leq1,  {\forall 1\leq i\leq N,}\label{cons:less1_x2}\\
   & \sum_{i=1}^{N}y'_{i,\bar{v},u,k}\leq k\cdot p_{\bar{v},u,k},{\forall \bar{v},u,k,}\label{cons:allocation_x2}\\
   & 0\leq y'_{i,\bar{v},u,k}\leq x_{i,\bar{v}}\cdot p_{\bar{v},u,k}\cdot\mathit{z_{i,\bar{v},u}},\label{cons:feasibility_x2}\\
   \mbox{var.} \;\;& y'_{i,\bar{v},u,k}\in [0,1], \mathbf{\mathbf{x}}_{i}\in\mathcal{X}_{i}, \nonumber \\
  & 1\leq i\leq N, 1\leq k\leq K, v\in\mathcal{V}^{[\tau]}, u\in\mathcal{V}. \nonumber 
\end{align}
Problem \textbf{MP-AN} is a linear-combinatorial one. The following corollary says that solving \textbf{MP-AN} also gives an $(1-1/e)$ approximation solution to \textbf{MP-T}.
\begin{mdframed}[skipabove=10pt, skipbelow=10pt, roundcorner=10pt, linewidth=0pt, backgroundcolor=gray!10] 
\begin{corol}
\label{cor:MP-AN}Let $\left(\boldsymbol{y}_{i}^{*},r_{i}^{*}\right)$,
$1\leq i\leq N$, be an optimal solution to \textbf{MP-T}, and let $\left(\bar{\boldsymbol{y}}_{i},\bar{\mathbf{x}}_{i}\right)$,
$1\leq i\leq N$, be that of \textbf{MP-AN}. Then 
\[
\left(1-1/e\right)\sum_{i=1}^{N}f_{i}\left(\boldsymbol{y}_{i}^{*},r_{i}^{*}\right)\leq\sum_{i=1}^{N}f_{i}\left(\bar{\boldsymbol{y}}_{i},\bar{r}_{i}\right)\leq\sum_{i=1}^{N}f_{i}\left(\boldsymbol{y}_{i}^{*},r_{i}^{*}\right),
\]
where $\bar{r}_i$ is the route for vehicle $i$ corresponding to $\bar{\mathbf{x}}_i$.
\end{corol}
\end{mdframed}

\textbf{Remark}: For problem \textbf{MP-AN}, one may think that we can use flow balance equations with unit flow from $s_i^0$ to $d_i^{-i}$ on $G^{[\tau]}$ and require the flow to take integer value to represent $\mathbf{x}_i\in \mathcal{X}_i$; relax the binary variables to $[0,1]$ and solve it as an LP. However, such relaxation in general incurs optimality loss, as we construct a counterexample with a positive integrity gap in 
\ifx \ISTR \undefined
our technical report \cite{TechnicalReport}.
\else
Appendix~\ref{app:integrity_gap}. 
\fi

\subsection{Dual Sub-Gradient Decent Algorithm}


We now present our dual sub-gradient algorithm for solving problem \textbf{MP-AN}. Relaxing the the right-hand-side constrains in \eqref{cons:feasibility_x2}  by introducing Lagrangian dual variables
\[
\mathbf{\lambda}\triangleq[\lambda_{i,\bar{v},u,k}]_{1\leq i \leq N,\bar{v}\in\mathcal{V}^{[\tau]},u\in\mathcal{V},1\leq k\leq K}\geq0,
\]
we obtain the following Lagrangian function,
\begin{align*}
 L\left(\mathbf{x},\mathbf{y},\mathbf{\lambda}\right) =& \sum_{i=1}^{N}\sum_{\bar{v}\in\mathcal{V}^{[\tau]}}\sum_{u\in\mathcal{V}}\sum_{k=1}^{K}\left(1-\lambda_{i,\bar{v},u,k}\right)y'_{i,\bar{v},u,k}\\
 & +\sum_{i=1}^{N}\sum_{\bar{v}\in\mathcal{V}^{[\tau]}}x_{i,\bar{v}}\sum_{u\in\mathcal{V}}\sum_{k=1}^{K}(\lambda_{i,\bar{v},u,k}\times p_{\bar{v},u,k}\times\mathit{z_{i,\bar{v},u}}).
\end{align*}
The dual function is then 
\begin{eqnarray*}
D(\mathbf{\lambda})= & \max & L(x,y,\lambda)\\
 & s.t. & (\ref{cons:less1_x2}), (\ref{cons:allocation_x2})\\
 &  \mbox{var.}& y'_{i,\bar{v},u,k}\in [0,1], \mathbf{\mathbf{x}}_{i}\in\mathcal{X}_{i}, \nonumber \\
 & & 1\leq i\leq N, 1\leq k\leq K, v\in\mathcal{V}^{[\tau]}, u\in\mathcal{V} .\nonumber
\end{eqnarray*}
As the variables $\mathbf{x}$ and $\mathbf{y}$ are decoupled in $D(\lambda)$, we can decompose $D(\lambda)$ into two optimization problems as follows:
\begin{eqnarray*}
\textbf{D1}: & \max & \sum_{i=1}^{N}\sum_{\bar{v}\in\mathcal{V}^{[\tau]}}\sum_{u\in\mathcal{V}}\sum_{k}^{K}\left(1-\lambda_{i,\bar{v},u,k}\right)y'_{i,\bar{v},u,k} \\
 & s.t. & (\ref{cons:less1_x2}), (\ref{cons:allocation_x2})\\
 & \mbox{var.} & y'_{i,\bar{v},u,k}\geq0, 1\leq i\leq N, 1\leq k\leq K, v\in\mathcal{V}^{[\tau]}, u\in\mathcal{V},\nonumber
\end{eqnarray*}
and
\begin{eqnarray*}
 \textbf{D2}: & \max & \sum_{i=1}^{N}\sum_{\bar{v}\in\mathcal{V}^{[\tau]}}x_{i,\bar{v}}\sum_{u\in\mathcal{V}}\sum_{k}^{K}(\lambda_{i,\bar{v},u,k}\times p_{\bar{v},u,k}\times\mathit{z_{i,\bar{v},u}})\\
 & \mbox{var.} & \mathbf{\mathbf{x}}_{i}\in\mathcal{X}_{i}, 1\leq i\leq N.\nonumber
\end{eqnarray*}
Note that the problem \textbf{D1} is simply an LP and can be solved with a complexity polynomial in the size of the regional network and maximum travel time deadline $\tau$. The problem \textbf{D2} is a longest path problem on an acyclic time-expanded regional graph, and it can be solved with a complexity polynomial in the size of the regional network and maximum travel time deadline $\tau$~\cite{Korte:2010:COT:1951946}. 

To this end, we arrive at an iterative dual-subgradient algorithm as follows: in each iteration, 
\begin{itemize}
    \item given a set of $\lambda_{i,\bar{v},u,k}$, we solve the problems \textbf{D1} and \textbf{D2} with a complexity polynomial in the size of the regional network and maximum travel time deadline $\tau$;
    \item we update the dual variables using a sub-gradient update: for $1\leq i \leq N,\bar{v}\in\mathcal{V}^{[\tau]},u\in\mathcal{V},1\leq k\leq K$,
        \[
            \lambda_{i,\bar{v},u,k}\leftarrow\lambda_{i,\bar{v},u,k}+\phi(\lambda)\left(y'_{i,\bar{v},u,k}-x_{i,\bar{v}}\times p_{\bar{v},u,k}\times\mathit{z_{i,\bar{v},u}}\right),
        \] 
        \rev{where $\phi(\lambda)$ is a diminishing step size \rev{ suggested} for subgradient algorithms~\cite{bazaraa1981choice}. The basic idea is choosing large initial stepsize and gradually decreasing the stepsize as the gap between the dual value and current recovered primal value gets smaller. And when we detect that the gap is smaller than a predefined threshold, we decrease the stepsize much faster. We leave the details to 
\ifx \ISTR \undefined
our technical report \cite{TechnicalReport}
\else
Appendix~To be add
\fi
.}

\end{itemize}
\rev{The dual-subgradient algorithm is known to converge at a rate of  $\mathcal{O}(\frac{1}{\sqrt{K}})$, where $K$ is the number of iteration. In our implementation, We terminate the iterations when either gap between the dual value and current recovered primal value gets smaller than a preset threshold or the number of iteration exceeds a preset limit.}



Upon convergence, the dual-subgradient algorithm is known to generate an optimal dual solution. However, it is not guaranteed to generate an optimal solution for the primal problem, since there could be duality gap for the linear-combinatorial problem studied in this section. In the following, we establish a condition under which our dual-subgradient algorithm also gives an optimal solution to the primal problem. 
\begin{mdframed}[skipabove=10pt, skipbelow=10pt, roundcorner=10pt, linewidth=0pt, backgroundcolor=gray!10] 
\begin{thm}
\label{thm:optimality} If upon termination of the dual-subgradient algorithm, the dual variables $\mathbf{\lambda}$ satisfy that
\begin{equation}
[y'_{i,\bar{v},u,k}-x_{i,\bar{v}}\times p_{\bar{v},u,k}\times\mathit{z_{i,\bar{v},u}}]_{\lambda_{i,\bar{v},u,k}}^{+}=0,\forall i,\bar{v},u,k,\label{eq:opt1}
\end{equation}
where function $[f]_{g}^{+}$ is defined as
\[
[f]_{g}^{+}=\begin{cases}
f, &\mbox{if } g>0;\\
\max(f,0), & \mbox{otherwise}.
\end{cases}
\]
Then each $x^{*}$ and \textup{$y^{*}$}, specifying routes for vehicles
and request-vehicle assignments, is an optimal solution to \textbf{MP-AN} and hence a $(1-1/e)$ approximation solution to \textbf{MP-T}.
\end{thm}
\end{mdframed}
The results are proved in 
\ifx \ISTR \undefined
our technical report \cite{TechnicalReport}
\else
Appendix~\ref{apx:proof_thm_optimality}
\fi
by utilizing an argument based on complementary slackness.

\textbf{Remarks.} We discuss how the dual-subgradient algorithm is implemented as follows. After attaining the solution $\mathbf{x}_i$ and $y'_{i,\bar{v},u,k}$, $\forall 1\leq i\leq N, 1\leq k\leq K, v\in\mathcal{V}^{[\tau]}, u\in\mathcal{V}.$  For vehicle $i$, it travels along the route $r_i$ to its destination. When vehicle $i$ passes by region $v_{i,j}$ in the corresponding slot $t_{i,j}$, given the actual requests information $u$ and $k$, we assign the request to the vehicles according to the request-vehicle assignment scheme in Sec.~\ref{ssec:joint.request-vehicle.assignment.and.routing}. If vehicle $i$ is assigned a request, then it follows a feasible ride-sharing plan by solving the feasibility problem in Sec.~\ref{ssec:ride-sharing.feasibility} to deliver both passengers (and such plan exists since the vehicle is assigned a request); otherwise, it goes to next region along the route $r_i$. 

%% file: simulation.tex
\begin{figure*}[t]
\subfigure[$\mathbb{P}\left(\sum_{d\in \mathcal{V}}\Phi_{s,t,d}\geq 1\right)$.\label{fig:probHeatLargerThan1}]{\includegraphics[width=0.45\columnwidth]{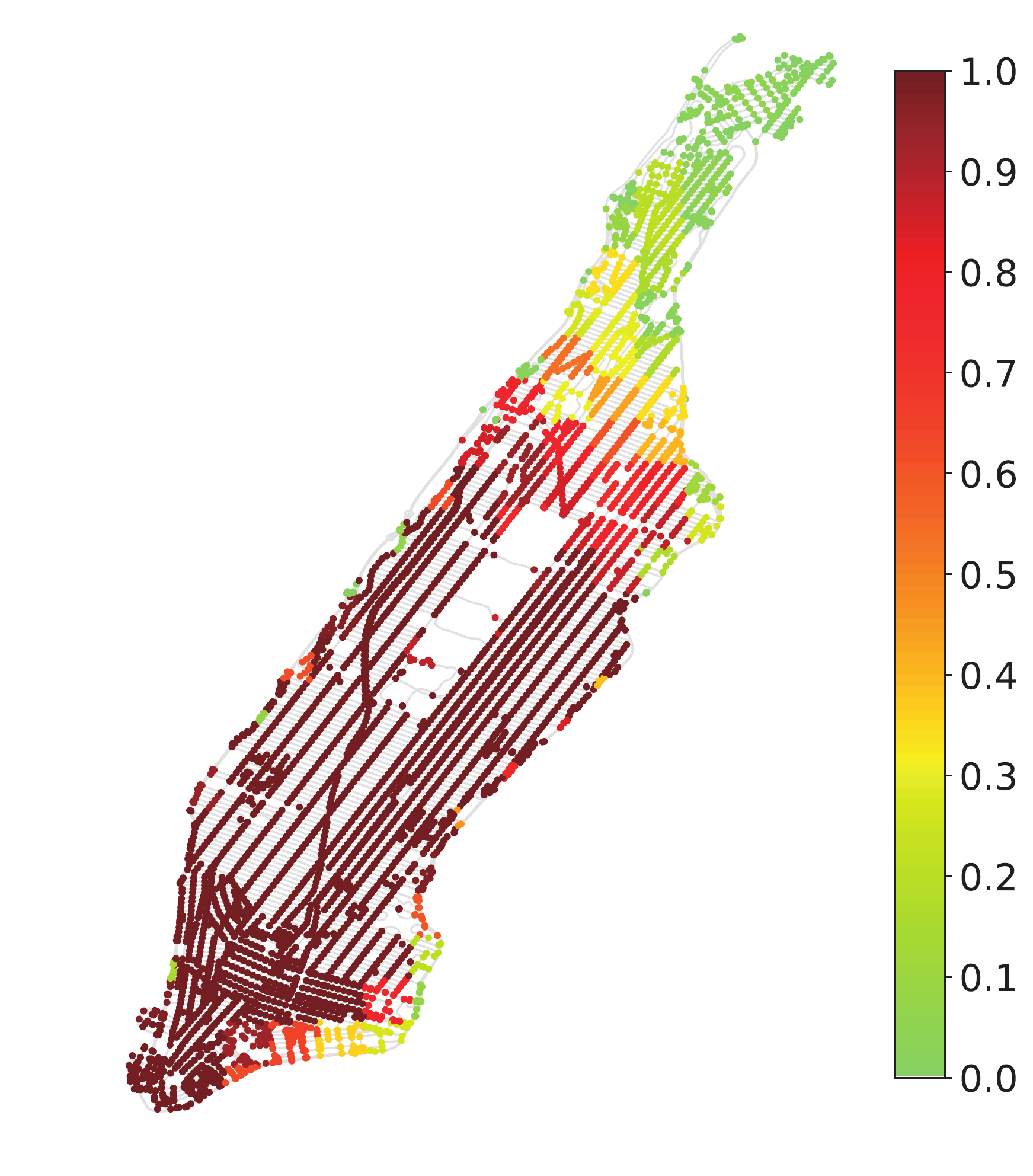}}
\subfigure[$\mathbb{P}\left(\sum_{d\in \mathcal{V}}\Phi_{s,t,d}\geq 10\right)$.\label{fig:probHeatLargerThan10}]{\includegraphics[width=0.45\columnwidth]{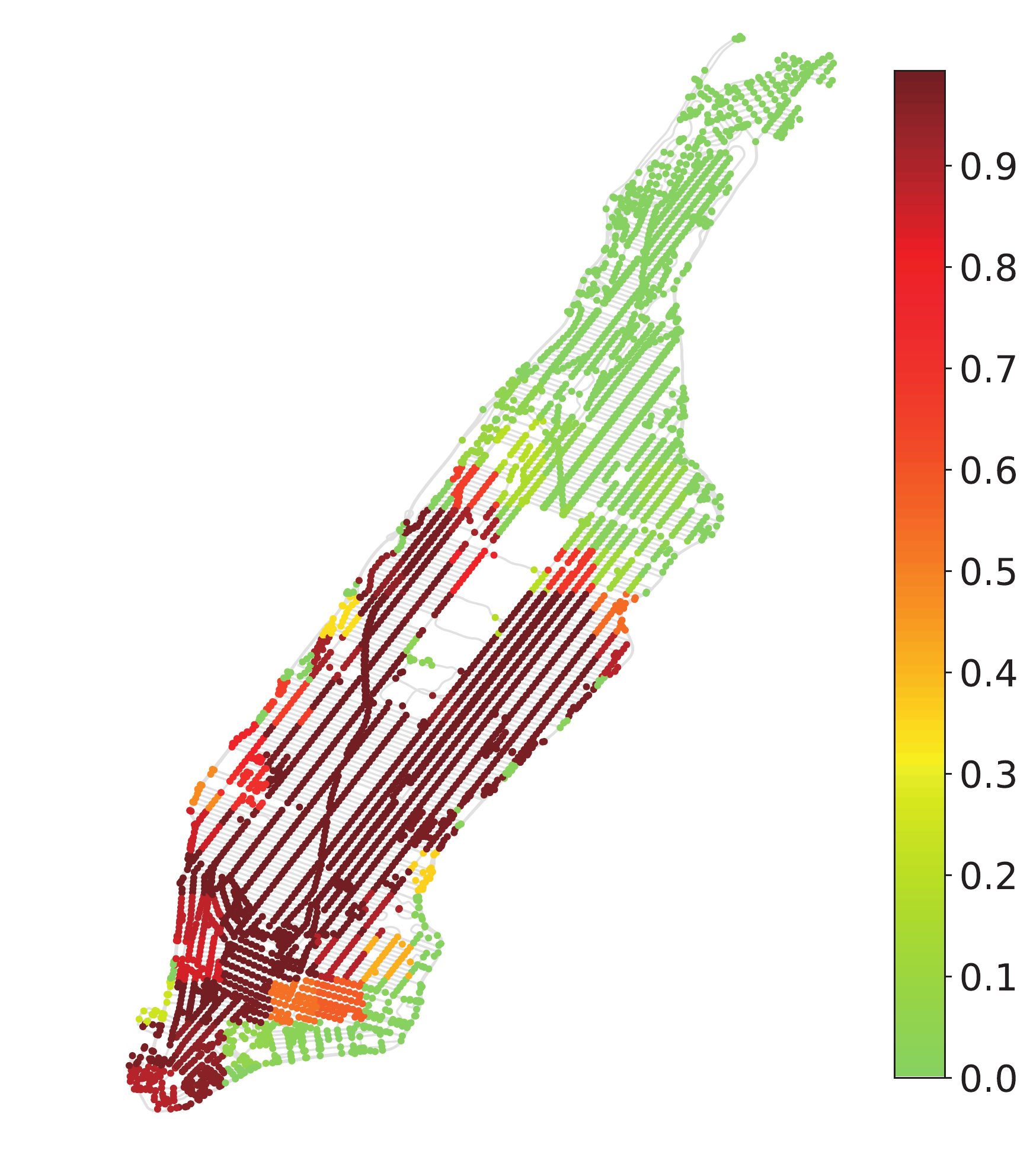}}
\subfigure[$\mathbb{P}\left(\sum_{d\in \mathcal{V}}\Phi_{s,t,d}\geq 20\right)$.\label{fig:probHeatLargerThan20}]{\includegraphics[width=0.45\columnwidth]{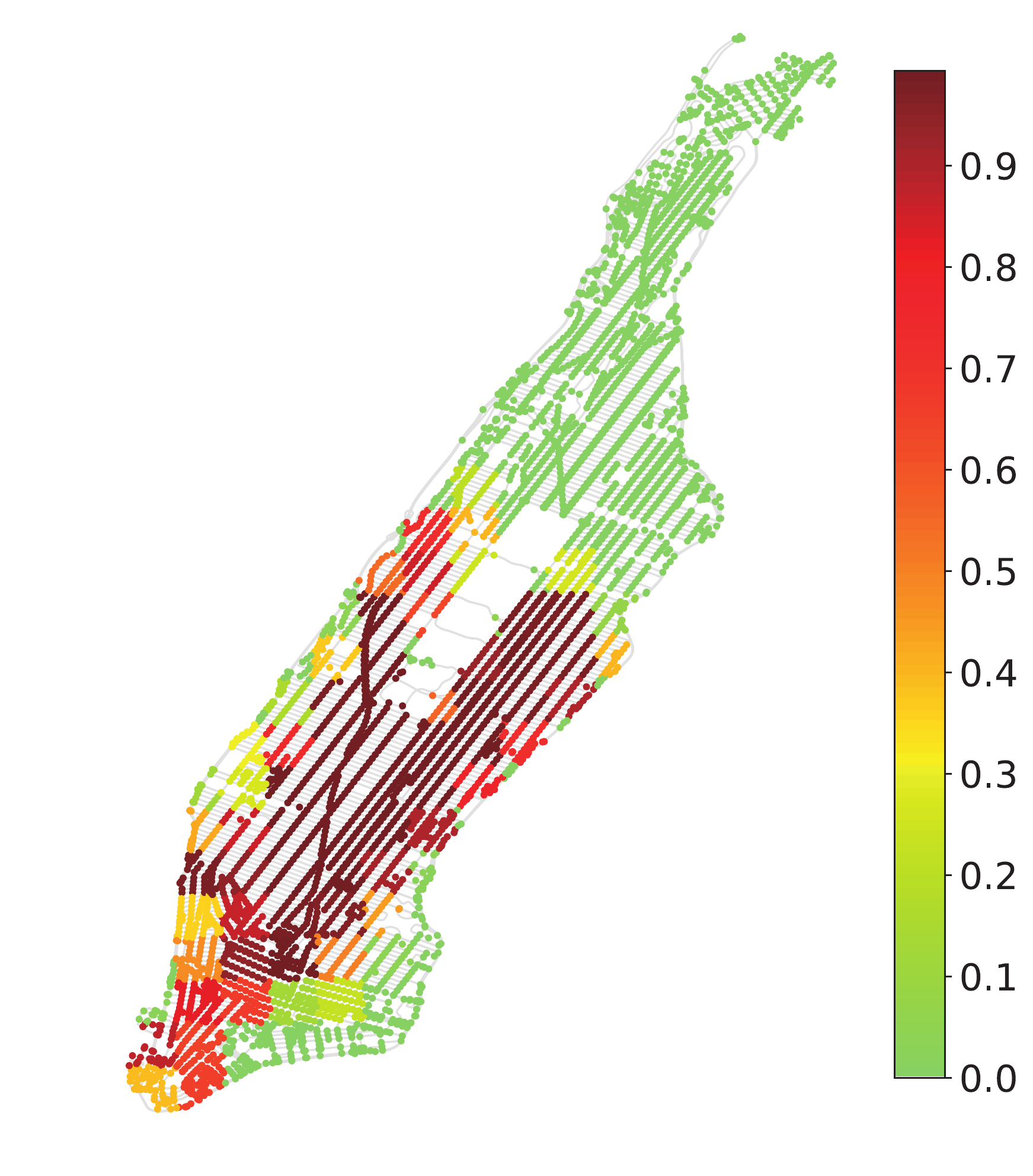}}
\subfigure[$\mathbb{P}\left(\sum_{d\in \mathcal{V}}\Phi_{s,t,d}\geq 30\right)$.\label{fig:probHeatLargerThan30}]{\includegraphics[width=0.45\columnwidth]{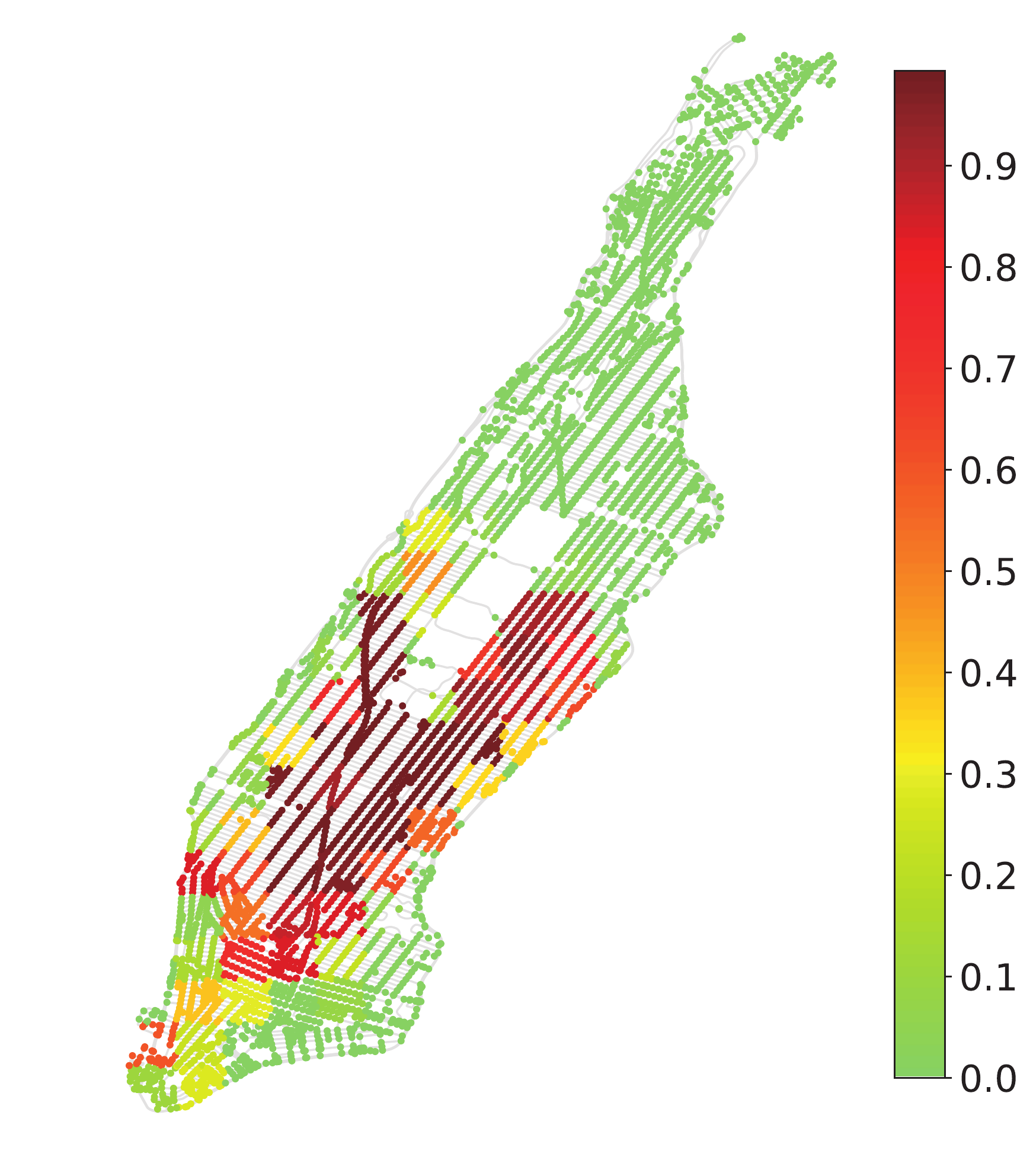}}
\caption{ \rev{Probability heatmaps calculated by using \eqref{equ:PrempCal} for the time slot $t=216$, corresponding to the time window $\left[18:00, 18:05\right]$ of the day. As seen, the distributions are highly heterogeneous across regions and many regions have limited demands. As such, it is important to perform a demand-aware, instead of demand-oblivious, design and to jointly assign requests to vehicles, in order to achieve proper demand-supply balancing and avoid passenger pickup conflict.}  
\label{fig:heatmap}
}
\end{figure*}

\section{{Numerical Experiments}\label{sec:simulation}}
In this section, we evaluate the performance of our solution of \rev{joint} request-vehicle assignment and vehicle routing for a fleet of vehicles, through extensive simulation using real-world \rev{traces}. \rev{Our purposes are to evaluate the benefit of (i) demand-aware ride-sharing at the fleet level as compared to the demand-oblivious baseline and (ii) our new joint fleet-level optimization as compared to the previous separate optimization \rev{at individual vehicles}. We use the number of fulfilled requests, \rev{terms as empirical pickups}, as the performance metrics. }



\subsection{Evaluation Setup}
\subsubsection{Dataset and Region Graph}
\par 
We \rev{use the taxi-trip dataset} of Manhattan, New York City~\cite{tripdata}. \rev{We extract around 60  million taxi-trip records in 6 months from the dataset (2016-01-01 to 2016-06-30), each including pickup time, pickup location, drop-off time, and drop-off location}. 
\par 
We obtain the Manhattan map from the OpenStreetMap~\cite{OpenStreetMap}. We then use python package NetworkX~\cite{Networkx} and OSMnx~\cite{OSMnx} to \rev{build a road network} $\mathcal{G}_0$.  Then we divide the whole area of 59.5$\text{km}^2$ into \rev{147 rectangular regions}, each of length 700 meters and \rev{width} 600 meters. \rev{We then construct
the region-based graph $\mathcal{G}$ according to the procedure described in Sec.~\ref{ssec:network-model}.}

We compute the distance $d_{u,v}$ (km) for edge $(u, v)$ in the \rev{ road network $\mathcal{G}_0$}. We assume that the taxis travel at the speed of 15km/h, according to a mobility report from NYC Department of Transportation~\cite{NYCReport}. \rev{The travel time of edge $(u, v)$ is estimated as}
\begin{equation*}
    \xi_{u,v} = \left\lceil \frac{60 \cdot d_{u,v}}{15} \right\rceil \text{ (minutes)}.
\end{equation*}

\subsubsection{Empirical distribution of travel \rev{requests}}
We use the trip data to \rev{obtain} the empirical distribution of the travel \rev{requests}. We first set the time slot length to be 5 minutes and divide the 24 hours in a day into \rev{$288$ slots}. The number of requests from  node $s$ to node $d$ \rev{at time slot $t\in[1,288]$ in a day} is modeled by a random variable $\Phi_{s,t,d}$. \rev{We use the taxi-trip data in the 6-month period, 182 days in total, to compute the empirical distribution as follows: for all $s$, $d$ in $\mathcal{G}$ and $t\in[1,288]$,
\begin{equation}
    \mathbb{P}\left(\Phi_{s,t,d}=k\right)=\frac{1}{182}\left(\mbox{\# days with $k$ $s-d$ requests in $t$}\right).
    \label{equ:PrempCal}
\end{equation}
We then plot the demand heat map to visualize the request distribution, i.e., $\mathbb{P}\left(\sum_{d\in V} \Phi_{s,t,d}\geq k\right)$, in Fig.~\ref{fig:heatmap}.}

\subsubsection{Simulation Environment}
We use a server cluster with 34 i7-3770/3.40GHz CPUs and 7 E5-2623/v3/3.00GHz CPUs \rev{for simulation}. Each machine has \rev{on average a} memory size of 17GB and has installed Red Hat as its operating system. The computing resource allows us to carry out real-world trace driven simulations for several hundreds of vehicles in the demand-crowded lower Manhattan area, consisting of ten 1.25 km x 1.25 km regions.\com{We should give a figure to illustrate which 10 regions that we pick for evaluation.} We use python to implement \rev{all the comparing algorithms}. We use python package Matplotlib to generate our figures.  

\subsubsection{Simulation Instance}
We use \rev{$\left(\Vec{s}, \Vec{d}, t_s\right)$} to denote a problem instance, where $\Vec{s}$ is the vector of sources of the N vehicles in the fleet, $\Vec{d}$ is the vector of destinations of the vehicles, and $t_s$ is the pickup time of the first passenger. In our simulation, we choose the first pickup time slot $t_s = t_i:=i\times\frac{60}{5}$, where $i=0,\dots,23$, in the $i-$th hour of one day. For every $t_s = t_i$, we  sample \rev{$10$} source and destination pairs \rev{with shortest path longer than 3} from the real world pickup traces in the particular hour $t_i$ in each of the 100 days that we run simulations upon. Once we have all the \rev{$10$} source destination pairs and the corresponding first-passenger pickup time $t_s$, we put them together to form one instance $(\Vec{s},\Vec{d},t_s)$. We generate in total more than $24\times 10=240$ instances for simulation. 
    
\subsubsection{Schemes for Comparison and Performance Metric}\label{sssec:schemes_for_comparison}
For each instance, we implement \rev{the following three algorithms and compare their performance}:
\begin{itemize}
    \item Joint routing: our demand-aware joint request-vehicle assignment and vehicle routing algorithm proposed in \rev{Sec.~\ref{sec:algorithm}}.
    \item Independent routing: our previous demand-aware routing algorithm for single vehicle only~\cite{QiulinRideSharingRouting17} and an intuitive uniform request-vehicle assignment scheme for assigning multiple appearing requests to multiple vehicles in the same region. 
    \item Fastest routing: \rev{a} demand-oblivious fastest routing algorithm and a uniform request-vehicle assignment scheme.
\end{itemize}
Note that here the uniform request-vehicle assignment scheme means that if there are more than one vehicle that can pick up a request in a region, we  assign the request to any of the vehicles uniformly at random. 

We evaluate an important performance metric \emph{empirical pickup}, namely the average total pickups of all the vehicles over 100 days. Intuitively, the empirical pickup represents the empirical \textit{service throughput} of the fleet in a region with certain demand distribution. \rev{Note that for each instance $(\Vec{s},\Vec{d},t_s)$, we evaluate the performance of a scheme by its average pickup across 100 days.}




\subsection{Benefits of Demand-Aware Optimization}
\rev{We evaluate the number of empirical pickups in different hours of a day of the three algorithms described in Sec.~\ref{sssec:schemes_for_comparison}.} In this evaluation, we set the number of vehicles to be N=50 and the delay tolerance factor to be $\alpha=1.3$ as defined in Sec.~\ref{ssec:demand-model} (used in our joint routing scheme and the independent routing scheme).

We recall that a request can be picked up by the fleet of N vehicles if and only if there is at least one vehicle that (i) is in the same \rev{5-min region as the request and (ii) is assigned by the request-vehicle assignment module to pick up the request.} If a request appears in a region and is not assigned to a vehicle, then the request cannot be fulfilled due to the maximum waiting time constraints. 

\begin{figure}[htbp]
    \centering
    \includegraphics[width=0.8\columnwidth]{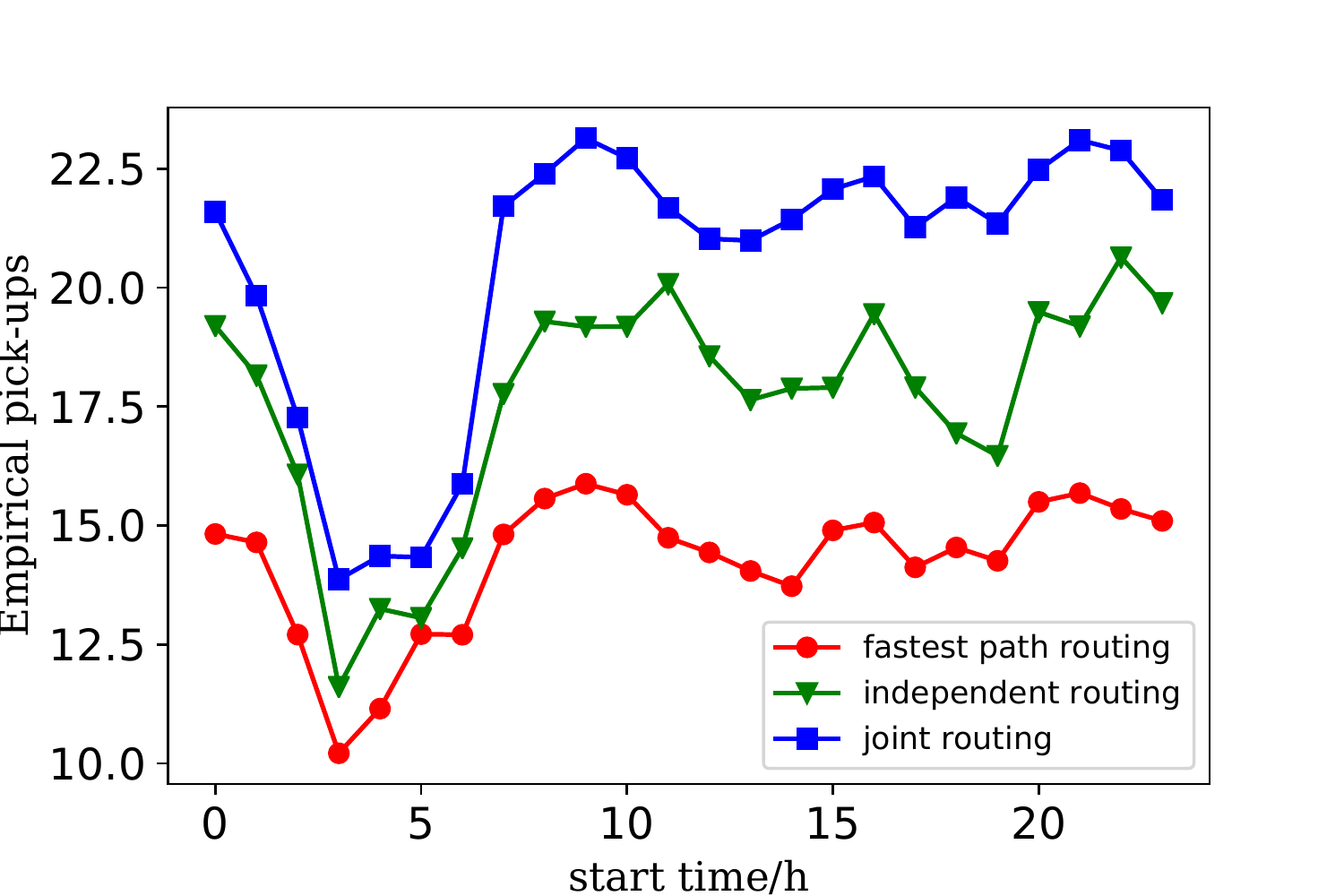}
    \caption{\rev{The empirical pickups under the setting of delay factor $\alpha=1.3$ and fleet size $N=50$}.}
    \label{fig:pickWithTime}
\end{figure}
The evaluation results are shown in \rev{Fig.~\ref{fig:pickWithTime}}. As seen, our proposed joint request-vehicle \rev{assignment} and routing algorithm fulfills significantly more requests than the independent routing algorithm and \rev{the} fastest path routing algorithm throughout the day, especially during the peak hours at \rev{noon and evening} (in Manhattan). \rev{In particular,} the daily-average improvement of our demand-aware solution as compared to the demand-oblivious fastest path routing is \rev{$46\%$}. This shows that exploiting demand statistics can significantly improve \rev{the service throughput of the fleet}. Furthermore, the daily-average improvement of our joint assignment and routing solution as compared to the independent routing solution is \rev{$19\%$}. \rev{This implies that joint optimization at the fleet-level can bring significant service throughput improvement as compared to the selfish optimization at the level of individual vehicles.}


\subsection{Impacts of the Fleet Size}
\begin{figure}[htbp]
    \centering
    \includegraphics[width=0.8\columnwidth]{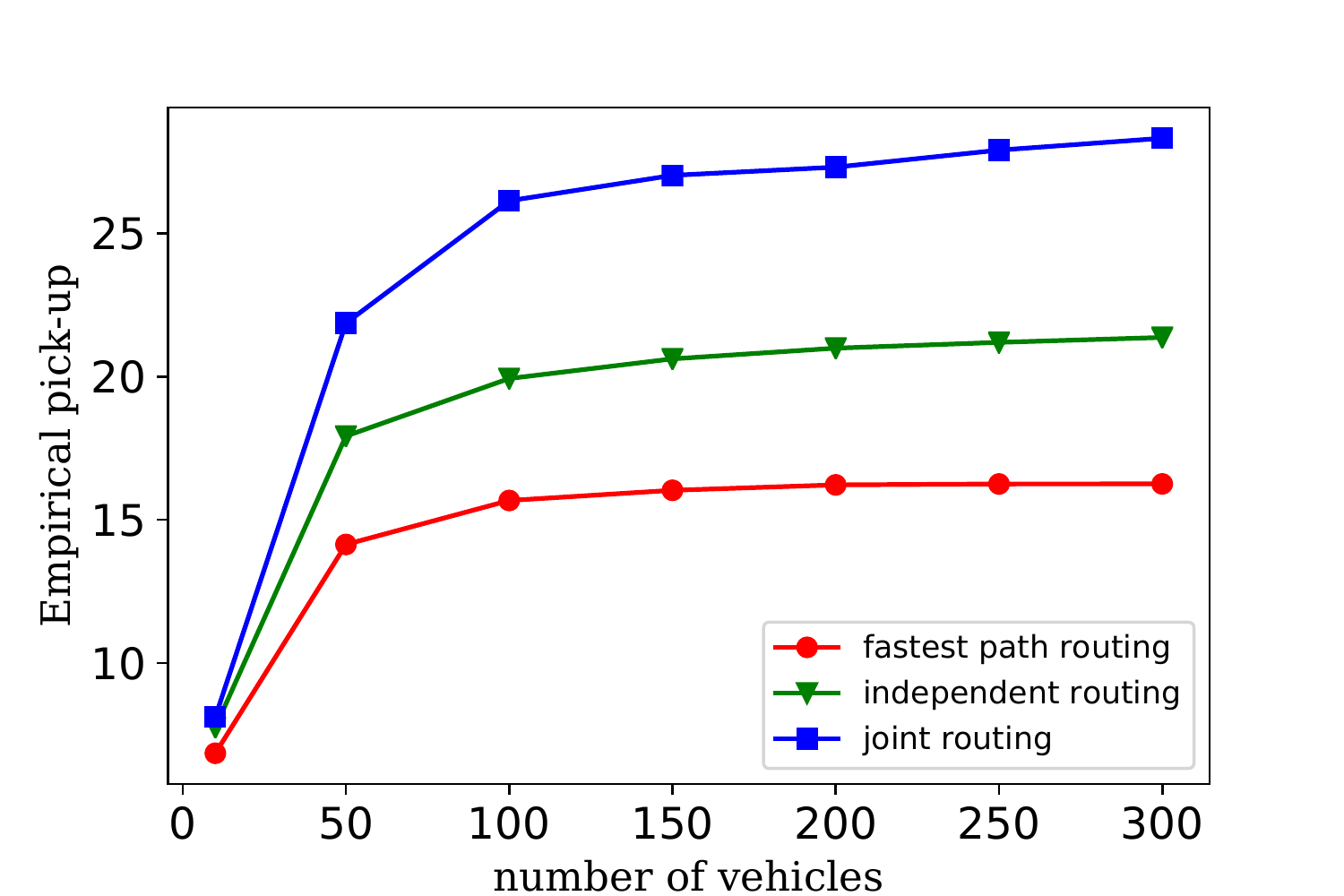}
    \caption{\rev{Empirical pickups for $\alpha=1.3$ and the time slot $t_s=204$, corresponding to the time window  $\left[17:00, 17:05\right]$ of the day.}}
    \label{fig:pickupChangingWithDriverNumber}
\end{figure}

We fix the delay tolerance factor to be $\alpha=1.3$ and the time slot \rev{to be $t_s = 204$,  corresponding to the time window  $\left[17:00, 17:05\right]$ of the day (when the request statistics is representative).} We evaluate how the performance of the three schemes vary as a function of the fleet size $N$. The results are reported in Fig.~\ref{fig:pickupChangingWithDriverNumber}. 

Ideally, \rev{for the demand-rich Manhattan area}, one would expect the number of pickups should increase as the fleet size $N$ increases. This is indeed the case for all the three \rev{algorithms when $N$ is less than 150}, as seen from \rev{Fig.~\ref{fig:pickupChangingWithDriverNumber}. }

Meanwhile, as the fleet size increases beyond 150 in our simulation, the number of pickups of the independent routing and fastest path routing saturate. In comparison, that of our joint routing solution still observes improvement. These two observations highlight two important insights. First, the saturation in service throughput improvements seen by the independent routing and the fastest path routing are not due to insufficient requests in the region. Rather, it is because neither of them considers \rev{load-balancing across regions}, which leads to inefficient routing decisions that result in excessive request-vehicle assignment conflicts in some regions while insufficient vehicles for serving requests in other regions. In contrast, our joint \rev{assignment and routing} solution properly load-balances the fleets (with request-fulfillment deadlines taken into consideration) across regions, allowing the service throughput of the fleet to \rev{increase as the fleet size increases.} 

Second, the results suggest that it is more important to perform intelligent fleet-level optimization for large fleets. This is also intuitive. \rev{When the fleet size is small, the limiting factor of the service throughput is the (small) number of vehicles.} In contrast, when the fleet size is large, the limiting factor is no longer the number of vehicles, but the routing and assignment efficiency. This explains the particularly superior performance of our joint \rev{assignment and routing} solution when the fleet size is large. Of course, when the fleet size further increases, one would expect that the limiting factor would change again to be the demand richness in the area, approaching the ``service capacity'' achievable by any fleets with optimal routing and assignment efficiency. 

Overall, this set of results suggest that it is important to perform fleet-level joint optimization to fully release the potential of demand-aware ride-sharing, in particular for large fleets.

\subsection{Impacts of the Delay Tolerance Factor}
\begin{figure}[htbp]
    \centering
    \includegraphics[width=0.8\columnwidth]{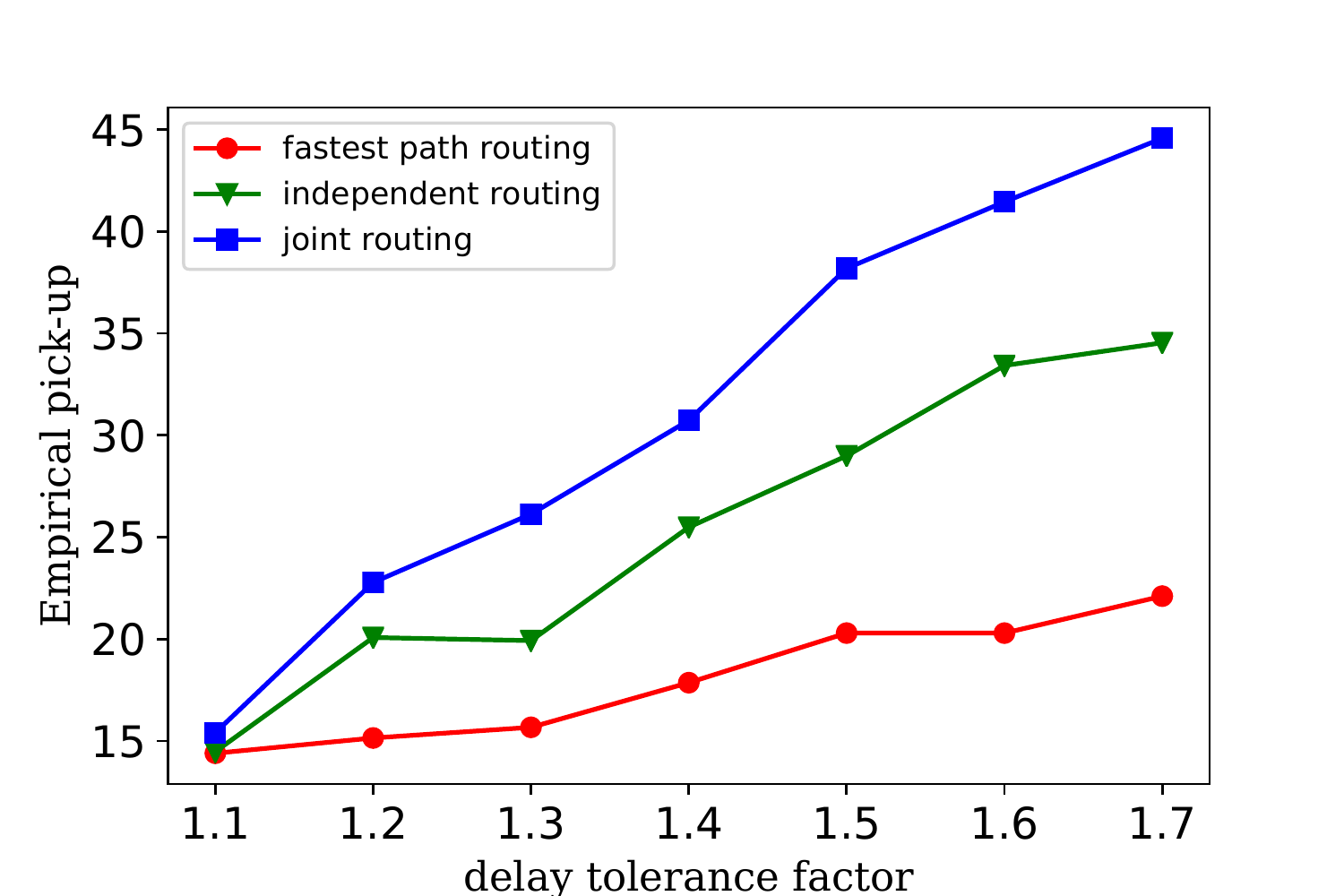}
    \caption{Empirical pickups for $N=100$ and the time \rev{slot  $t_s=204$, corresponding to the time window  $\left[17:00, 17:05\right]$ of the day}.}
    \label{fig:pickupChangingWithDelayFactor}
\end{figure}

To study the impacts of delay tolerance factor $\alpha$, we fix \rev{$t_s = 204$,  corresponding to the time window  $\left[17:00, 17:05\right]$ of the day. We plot the empirical pickups as a function of the delay factor in Fig.~\ref{fig:pickupChangingWithDelayFactor}.}
Again, our proposed joint assignment and routing algorithm outperforms the other two significantly. We observe that when the delay factor $\alpha$ increases, the empirical pickups also increase, which is intuitive as the \rev{ride-sharing} optimization space increases as $\alpha$ increases. \rev{We also observe that the pickups of the demand-oblivious fastest path routing algorithm increases slower than the demand-aware solutions. }

\rev{Furthermore, we observe that the performance gap between our proposed joint assignment and routing algorithm and the independent routing algorithm increase as the delay factor increases. An intuitive explanation is that as the delay factor increases, the ride-sharing optimization spaces for individual vehicles increase. Thus it becomes more likely that different vehicles decide to route through the same demand-rich regions, causing more pickup conflicts and serious service imbalance across regions. On the contrary, our joint assignment and routing approach properly load-balances across the regions and thus maximizes the empirical pickups.} 


%% file: conclusion.tex
\section{Concluding Remarks}\label{sec:conclusion}
\rev{
As the first step to explore the demand-aware design, this paper focuses on the snapshot
version of the ride-sharing problem. While the problem is already NP-hard, we derive a practical pseudo polynomial-time algorithm that achieves an approximation ratio of $(1-1/e)$ under the conditions presented in Theorem~\ref{thm:optimality}. 

We make the following remarks. First, in our joint routing and request-vehicle assignment optimization at the present time, i.e., $t=0$, the statistical future demand information at $t=1,2...$ is already taken into account. Thus our approach is a demand-aware one for the \rev{snapshot} version of the ride-sharing problem.  Second, upon change in vehicle status, e.g., a user is delivered to the destination or an empty car picks up a new user, the system can re-optimize the routing decisions and request-vehicle assignments, so as to optimize the long-term ride-sharing performance in a greedy fashion. The
overall solution can serve as a baseline for other demand-aware studies, e.g., the conceivable ones by extending the
approach in~\cite{EmptyCar} and~\cite{oda2018movi} to the multi-rider setting and the recent one in~\cite{alabbasi2019deeppool}. 
We leave the performance analysis of the overall greedy solution, as well as developing solutions with optimized long-term performance, as
interesting and important future directions.
}

%% file: Appendix.tex
\section{Appendix}

\subsection{A Counterexample on Integrity Gap}
\label{app:integrity_gap}
\begin{figure}[ht]
\includegraphics[width=0.6\columnwidth]{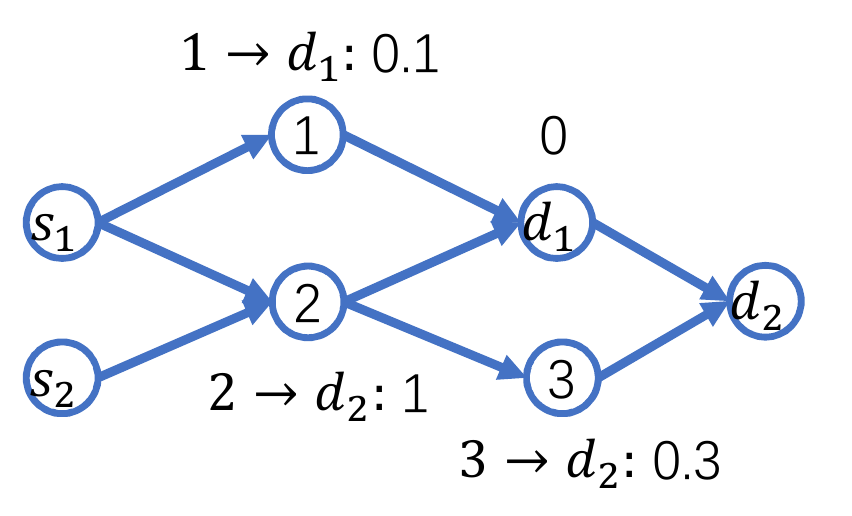}
\caption{An example with positive integrity gap. \label{fig:integrity_gap}}
\end{figure}

We provide an example where the integrity gap of problem \textbf{MP-N2} is {positive} in Fig. ~\ref{fig:integrity_gap}. On the transportation network, the travel time of each edge is {one} slot. We assume the delay factor $\alpha=1$. The future request statistics is as follows: {
\begin{itemize}
    \item $1\to d_1$: at node $1$, slot $1$, there is 1 request to $d_1$ with probability 0.1;
    \item $2\to d_2$: at node $2$, slot $1$, there is 1 request to $d_2$ with probability 1;
    \item $3\to d_2$: at node 3, at slot $2$, there is 1 request to $d_2$ with probability 0.3. 
\end{itemize}}
Suppose there are two vehicles. Vehicle 1 picks up a passenger from node $s_1$ to $d_2$. Vehicle 2 picks up a passenger from node $s_2$ to $d_2$. Then {an optimal integer solution} is that vehicle 1 travels along $s_1\to 2 \to d_1$ and is allocated a probability of 0.3 to pick up a request $2 \to d_2$, while vehicle 2 travels along $s_2 \to 2 \to 3 \to d_2$ and is allocated a probability of 0.7 to pick up a request $2\to d_2$ and a probability of 0.3 to pick up request $3\to d_2$. The optimal value is then 1.3. 

However, we can construct a {fractional} solution with larger objective value. Vehicle 2 remains the same, but vehicle 1 travels along $s_1\to 2\to d_1$ with a flow portion of 0.3 (i.e., $x_{1,2}=0.3$) and along $s_1 \to 1\to d_1$ with a flow portion of 0.7 (i.e $x_{1,1}=0.7$) and is allocated a probability of 0.3 to pick up a request $2\to d_2$, a probability of 0.07 to pick up a request $1\to d_1$ (satisfying (\ref{cons:feasibility_x2})). {The objective value is 1.37 then which is large than the value 1.3 obtained by the optimal integer solution.}  

\subsection{Prove of Theorem~\ref{thm:MP-AN}.}
\label{app:them:MP-AN}
\begin{proof}
We prove it by showing that \textbf{MP-AN} is equivalent to \textbf{MP-A}. Then following Theorem~\ref{thm:approx}, we easily conclude it. We claim that each feasible solution of \textbf{MP-A} can be mapped to a feasible solution of \textbf{MP-AN} with the same objective value and vice verse. 

First, suppose there is a feasible solution $\triangleq(\mathbf{y}_i),r_i)$ of \textbf{MP-A}, we then construct a solution $\triangleq(\bar{\mathbf{y}}_i,\bar{\mathbf{x}}_i)$ of \textbf{MP-AN} in the following way and show the two solutions share the same objective values. Given $r_i$, $\bar{x}_i$ is constructed according to (\ref{eq:route2node}). Define $\tilde{\mathbf{y}}_i=\left[\tilde{y}_{i,\bar{v},u,k},\;\forall v\in\mathcal{V}^{[\tau]},u\in\mathcal{V},1\leq k\leq N\right]$, where 
\[
\tilde{y}_{i,\bar{v},u,k}=\bar{x}_{i,\bar{v}}\times y_{i,\bar{v},u,k},
\]
and,
\[
h_{i}\left(\mathbf{y_{i}},r_{i}\right)\triangleq\max\left(1,\sum_{j=1}^{n_{i}}\sum_{u\in\mathcal{V}}\sum_{k=1}^{K}y_{i,\bar{v}_{i,j},u,k}\right).
\]
Then set $\bar{\mathbf{y}}_i=\left[\bar{y}_{i,\bar{v},u,k},\;\forall v\in\mathcal{V}^{[\tau]},u\in\mathcal{V},1\leq k\leq N\right]$, where
\[
\bar{y}_{i,\bar{v},u,k}=\frac{\tilde{y}_{i,\bar{v},u,k}}{h_{i}\left(\mathbf{y_{i}},r_{i}\right)}.
\]
We can easily check constraints (\ref{cons:less1_x2}), (\ref{cons:allocation_x2}) and (\ref{cons:feasibility_x2}) to ensure the feasibility of $(\bar{\mathbf{y}}_i,\bar{\mathbf{x}}_i)$. The result that the objective values are the same follows the fact that 
\begin{align*}
    g_{i}\left(\mathbf{y_{i}},r_{i}\right) & = \min\left(1,\sum_{j=1}^{n_{i}}\sum_{u\in\mathcal{V}}\sum_{k=1}^{K}y_{i,\bar{v}_{i,j},u,k}\right)\\
    & =\frac{\sum_{\bar{v}\in\mathcal{V}^{[\tau]}}\sum_{u\in\mathcal{V}}\sum_{k=1}^K \bar{x}_{i,\bar{v}}y_{i,\bar{v},u,k}}{h_{i}\left(\mathbf{y_{i}},r_{i}\right)}
\end{align*}

Second, suppose there is a feasible solution $\triangleq(\bar{\mathbf{y}}_i,{\mathbf{x}}_i)$ of \textbf{MP-AN}, we then construct a solution $\triangleq(\mathbf{y}_i),r_i)$ of \textbf{MP-A} in the following way and show the two solutions share the same objective values. $r_i$ is the route in the region graph corresponding to ${\mathbf{x}}_i)$. Note that as the time-expanded graph is acyclic and directed, $\mathbf{x}_i$ determines a unique route. We then construct $\mathbf{y}_i$ in the as follow. ${\mathbf{y}}_i=\left[{y}_{i,\bar{v},u,k},\;\forall v\in\mathcal{V}^{[\tau]},u\in\mathcal{V},1\leq k\leq N\right]$, where
\[
{y}_{i,\bar{v},u,k}={\bar{x}_{i,\bar{v}}}\times{\bar{y}_{i,\bar{v},u,k}}.
\] 
Also, we can easily check that $(\mathbf{y}_i),r_i)$ satisfies the constraints in \textbf{MP-A} and thus it is feasible. We show that they have the same objective value by 
\begin{align*}
     g_{i}\left(\mathbf{y_{i}},r_{i}\right) & = \min\left(1,\sum_{j=1}^{n_{i}}\sum_{u\in\mathcal{V}}\sum_{k=1}^{K}y_{i,\bar{v}_{i,j},u,k}\right)\\
    & =\min\left(1,\sum_{j=1}^{n_{i}}\sum_{u\in\mathcal{V}}\sum_{k=1}^{K}{\bar{x}_{i,\bar{v}_{i,j}}}\times{\bar{y}_{i,\bar{v}_{i,j},u,k}}\right)\\
    & =\min\left(1,\sum_{\bar{v}\in\mathcal{V}^{[\tau]}}\sum_{u\in\mathcal{V}}\sum_{k=1}^{K}{\bar{x}_{i,\bar{v}}}\times{\bar{y}_{i,\bar{v},u,k}}\right)\\
    & \stackrel{(a)}{=} \min\left(1,\sum_{\bar{v}\in\mathcal{V}^{[\tau]}}\sum_{u\in\mathcal{V}}\sum_{k=1}^{K}{\bar{y}_{i,\bar{v},u,k}}\right)\\
    & = \sum_{\bar{v}\in\mathcal{V}^{[\tau]}}\sum_{u\in\mathcal{V}}\sum_{k=1}^{K}{\bar{y}_{i,\bar{v},u,k}}
\end{align*}
, where (a) is by the fact that $\bar{y}_{i,\bar{v},u,k}={\bar{x}_{i,\bar{v}}}\times{\bar{y}_{i,\bar{v},u,k}}$ ( 1) if $\bar{x}_{i,\bar{v}}=1$, it holds; 2)if $\bar{x}_{i,\bar{v}}=0$, then following (\ref{cons:feasibility_x2}), $\bar{y}_{i,\bar{v},u,k}=0$ and thus it holds). 

We then conclude that \textbf{MP-AN} is equivalent to \textbf{MP-A}. Then following Theorem~\ref{thm:approx}, we have Theorem~\ref{thm:MP-AN} straightly.
\end{proof}


\subsection{Prove of Theorem~\ref{thm:optimality}}\label{apx:proof_thm_optimality}
\begin{proof}
First, we show that the solution $(\mathbf{y}'_i, \mathbf{x}_i)$ satisfying {the condition in \eqref{eq:opt1}} is a feasible solution to \textbf{MP-AN}. To see this, {we only need to check if the following constraint is satisfied} 
\begin{equation}
\label{cons:relaxing}
y'_{i,\bar{v},u,k}\leq x_{i,\bar{v}}\cdot p_{\bar{v},u,k}\cdot\mathit{z_{i,\bar{v},u}}.
\end{equation}
If the corresponding $\lambda_{i,\bar{v},u,k}$ is strictly positive, {the condition in} \eqref{eq:opt1} implies that
\[
y'_{i,\bar{v},u,k}-x_{i,\bar{v}}\times p_{\bar{v},u,k}\times\mathit{z_{i,\bar{v},u}}=0,
\]
and thus $(\mathbf{y}'_i, \mathbf{x}_i)$ satisfies \eqref{cons:relaxing}. 

{Otherwise the corresponding $\lambda_{i,\bar{v},u,k}=0$. Based on the condition in \eqref{eq:opt1},} we have that 
\[
    \max\left(y'_{i,\bar{v},u,k}-x_{i,\bar{v}}\times p_{\bar{v},u,k}\times\mathit{z_{i,\bar{v},u}},0\right
    )=0,
\]
{and consequently}
\begin{equation}
\label{ieq:lower}
    y'_{i,\bar{v},u,k}-x_{i,\bar{v}}\times p_{\bar{v},u,k}\times\mathit{z_{i,\bar{v},u}}\leq 0. 
\end{equation}

We then conclude that $\left(\mathbf{y}'_i, \mathbf{x}_i\right)$ satisfies \eqref{cons:relaxing}. Overall, the solution is feasible and hence provides a lower bound for the optimal value of the problem \textbf{MP-AN}, {denoted as $OPT$}, i.e.,
\[
\sum_{i=1}^{N}\sum_{\bar{v}\in\mathcal{V}^{[\tau]}}\left(\sum_{u\in\mathcal{V}}\sum_{k=1}^{K}y'_{i,\bar{v},u,k}\right)\leq OPT.
\]

Second, we show its {optimality}. We look at the dual function value of the solution $(\mathbf{y}'_i, \mathbf{x}_i)$. 
\begin{align*}
    D(\mathbf{\lambda}) &= \sum_{i=1}^{N}\sum_{\bar{v}\in\mathcal{V}^{[\tau]}}\sum_{u\in\mathcal{V}}\sum_{k=1}^{K}\left(1-\lambda_{i,\bar{v},u,k}\right)y'_{i,\bar{v},u,k}\\
    & +\sum_{i=1}^{N}\sum_{\bar{v}\in\mathcal{V}^{[\tau]}}x_{i,\bar{v}}\sum_{u\in\mathcal{V}}\sum_{k=1}^{K}(\lambda_{i,\bar{v},u,k}\times p_{\bar{v},u,k}\times\mathit{z_{i,\bar{v},u}})\\
    & = \sum_{i=1}^{N}\sum_{\bar{v}\in\mathcal{V}^{[\tau]}}\sum_{u\in\mathcal{V}}\sum_{k=1}^{K}y'_{i,\bar{v},u,k}\\
    & -\sum_{i=1}^{N}\sum_{\bar{v}\in\mathcal{V}^{[\tau]}}\sum_{u\in\mathcal{V}}\sum_{k=1}^{K}\lambda_{i,\bar{v},u,k}\times\left( y'_{i,\bar{v},u,k}-x_{i,\bar{v}} p_{\bar{v},u,k}\times\mathit{z_{i,\bar{v},u}}\right)\\
    & = \sum_{i=1}^{N}\sum_{\bar{v}\in\mathcal{V}^{[\tau]}}\sum_{u\in\mathcal{V}}\sum_{k=1}^{K}y'_{i,\bar{v},u,k}.
\end{align*}
The last equality is {by the condition in \eqref{eq:opt1}, which implies}
\[
\left( y'_{i,\bar{v},u,k}-x_{i,\bar{v}} p_{\bar{v},u,k}\times\mathit{z_{i,\bar{v},u}}\right)=0.
\]
Then by weak duality we have 
\begin{equation}
\label{ieq:upper}
D(\mathbf{\lambda}) = \sum_{i=1}^{N}\sum_{\bar{v}\in\mathcal{V}^{[\tau]}}\sum_{u\in\mathcal{V}}\sum_{k=1}^{K}y'_{i,\bar{v},u,k}\geq OPT.
\end{equation}
{By \eqref{ieq:lower} and \eqref{ieq:upper}, $\left(\mathbf{y}'_i, \mathbf{x}_i\right)$ satisfying \eqref{eq:opt1} is an optimal solution to problem \textbf{MP-AN}}.
\end{proof}